\documentclass[conference]{IEEEtran}
\IEEEoverridecommandlockouts
\usepackage{cite}
\usepackage{textcomp}

\usepackage{microtype}

\usepackage{epsfig,amssymb,amsmath}
\usepackage{amsfonts}
\usepackage{graphicx,color}
\usepackage{bm}
\usepackage{url}

\usepackage{multirow}

\usepackage{algorithm}
\usepackage{algpseudocode}
\usepackage{algpascal}

\allowdisplaybreaks

\newtheorem{fact}{Fact}
\newtheorem{theorem}{Theorem}[section]
\newtheorem{proposition}[theorem]{Proposition}
\newtheorem{corollary}[theorem]{Corollary}
\newtheorem{lemma}[theorem]{Lemma}

\def\poa{{\sf PoA}}
\def\pos{{\sf PoS}}

\begin{document}

\title{On Colorful Bin Packing Games}
\author{
\IEEEauthorblockN{Vittorio Bil\`o}
\IEEEauthorblockA{
\textit{Dept of Mathematics and Physics,} \\
\textit{University of Salento}\\
Italy \\
vittorio.bilo@unisalento.it}
\and
\IEEEauthorblockN{Francesco Cellinese}
\IEEEauthorblockA{
\textit{Gran Sasso Science Institute} \\
Italy  \\
francesco.cellinese@gssi.it}
\and
\IEEEauthorblockN{Giovanna Melideo}
\IEEEauthorblockA{\textit{DISIM} \\
\textit{University of L'Aquila}\\
Italy \\
giovanna.melideo@univaq.it}
\and
\IEEEauthorblockN{Gianpiero Monaco}
\IEEEauthorblockA{\textit{DISIM} \\
\textit{University of L'Aquila}\\
Italy \\
gianpiero.monaco@univaq.it}
}

\maketitle

\begin{abstract}
We consider colorful bin packing games in which selfish players control a set of items which are to be packed into a minimum number of unit capacity bins. Each item has one of
$m\geq 2$ colors and cannot be packed next to an item of the same color. All bins have the same unitary cost which is shared among the items it contains, so that players are interested in selecting a bin of minimum shared cost. We adopt two standard cost sharing functions: the egalitarian cost function which equally shares the cost of a bin among the items it contains, and the proportional cost function which shares the cost of a bin among the items it contains proportionally to their sizes.
Although, under both cost functions, colorful bin packing games do not converge in general to a (pure) Nash equilibrium, we show that Nash equilibria are guaranteed to exist and
we design an algorithm for computing a Nash equilibrium whose running time is polynomial under the egalitarian cost function and pseudo-polynomial for a constant number of colors under the proportional one.
We also provide a complete characterization of the efficiency of Nash equilibria under both cost functions for general games, by showing that the prices of anarchy and stability are unbounded when $m\geq 3$ while they are equal to 3 for black and white games, where $m=2$.
We finally focus on games with uniform sizes (i.e., all items have the same size) for which the two cost functions coincide.
We show again a tight characterization of the efficiency of Nash equilibria and design
an algorithm which returns Nash equilibria with best achievable performance.
\end{abstract}


\section{Introduction}
A classical problem in combinatorial optimization is the one-dimensional bin packing
problem, in which items with different sizes in $[0,1]$ have to be packed
into the smallest possible number of unit capacity bins.
This problem is known to be NP-hard (see \cite{CGJ96} for a survey).

The study of bin packing in a game theoretical context has been introduced in \cite{B06}. In such a setting, items are handled by selfish players and the unitary cost of each bin is shared among the items it contains. In the literature, two natural cost sharing functions have been considered. The {\em egalitarian} cost function, used in \cite{DE14,MDHTYZ13}, which equally shares the cost of a bin among the items it contains, and the {\em proportional} cost function, studied in \cite{B06,DE12}, where the cost of a bin is split among the items proportionally to their sizes. That is, each player is charged with a cost according to the fraction of the used bin space her item requires.
We notice that for games with uniform sizes, i.e., when all the items have the same size $s$, the two cost functions coincide.
Each player would prefer to choose a strategy that minimizes her own cost, where the strategy is the bin chosen by the player.
Pure Nash equilibria, i.e. packings in which no player can lower her cost by changing the selected bin in favor of a different one, are mainly considered as natural stable outcomes for these games.
The social cost function that we aim to minimize is the number
of open bins (a bin is open if it stores at least one item).
Bin packing games can be used to model many practical scenarios, like for instance, bandwidth allocations problems, packet scheduling problems, etc. (see \cite{B06,EK11}).

In this paper we consider {\em colorful bin packing} games, where we are given a set of $n$ selfish players, a set of $m \geq 2$ colors, and a set of $n$ unit capacity bins.
Players control indivisible colored items of size in $[0,1]$.
Each item needs to be packed into a bin without exceeding its capacity and in such a way that no item is misplaced, that is no item is packed next to an item of the same color.
The special case of games with two colors is called {\em black and white} bin packing games. We use both the egalitarian and the proportional cost functions, where we set the cost of any misplaced item as infinite, and adopt Nash equilibria as stable outcomes of the games. Clearly, in any Nash equilibrium no player can be charged with a infinite cost, since, for any player, moving a misplaced item to an empty bin (and thus getting cost $1$) is always an improving deviation. We notice that, if all the items have different colors, these games correspond to the bin packing ones. However, when there are items with the same color, the stable outcomes of the games are structurally different than bin packing ones. In fact, we show that in our games, Nash equilibria perform very differently than in bin packing ones.

Colorful bin packing games generalize the bin packing games and therefore model many practical scenarios and situations where it is important distinguishing or separating the items, that are not captured by bin packing games. For instance, suppose that a shipping company rents out its means of transport like trucks. Customers want to ship their items at the smallest cost. For example, items could be chemical agents with different characteristics or properties such that, if two items with the same property are packed next to the other in the truck, they could dangerously react. Colorful bin packing games can be used to model such scenario. Items here correspond to chemical agents, colors to characteristics, and bins to trucks.
Other interesting applications can be found in \cite{BDESV,DE14b}.

\noindent{\bf Our Contribution.}
We first focus on the existence of Nash equilibria. We show that colorful bin packing games may not converge to Nash equilibria. In fact, they may admit an infinite sequence of improving deviations (i.e., the finite improvement path property is not guaranteed) even for special cases in which games have only two colors and uniform sizes (Proposition \ref{prop1}). In this case the egalitarian and proportional cost functions coincide.
However, in Theorems \ref{thm1} and \ref{thm2}, we show that, under both cost functions, if one allows the players to perform only improving deviations towards bins in which no item is misplaced, then any game possesses the finite improving path property. As in any non-equilibrium profile there always exists a player who possesses one such a deviation, and in particular a misplaced item can always move to an empty bin, it follows that Nash equilibria are guaranteed to exist under both cost functions. We also show a very natural and simple algorithm Algorithm \ref{alg:ne} (a similar approach was already considered in \cite{DE12}), that computes a Nash equilibrium whose running time is polynomial under the egalitarian cost function and pseudo-polynomial for a constant number of colors under the proportional one (Theorem \ref{thm:computing_equilibria_both_cost_function}).

We then measure the quality of Nash equilibria using the standard notions of $\poa$ (price of anarchy) and $\pos$ (price of stability), that are defined as the worst/best case ratio between the social cost of a Nash equilibrium and the cost of a social optimum, which corresponds to the minimum number of open bin needed to feasibly pack all colored items. We provide a complete characterization of the efficiency of Nash equilibria by showing that, under both cost functions, the prices of anarchy and stability of colorful bin packing games are unbounded (we consider the absolute approximation ratio), when $m\geq 3$ (Theorems \ref{POS_unbounded_egalitarian_m>=3} and \ref{POS_unbounded_proportional_m>=3}), while they are equal to $3$ when $m=2$ (Theorems \ref{ubpoa}, \ref{thm:lowerbound_pos_twocolors_egalitarian}, and \ref{thm:lowerbound_pos_twocolors_proportional}). We also consider the basic setting in which all items have the same size $s$ and again provide a complete picture of the efficiency of Nash equilibria which happens to depend on the parity of the number $\kappa=\lfloor 1/s\rfloor$ of items that can be packed into a bin without exceeding its capacity (in this case the egalitarian and proportional cost functions coincide). In particular, we show that, when $\kappa$ is even, the price of stability is $2$ for any
$m\geq 2$ (Theorems \ref{lbposuniform} and \ref{ubposuniform}), while the price of anarchy is $2$ for $m=2$ (Theorem \ref{upperbound_poa_twocolors_k_even}), and unbounded for $m\geq 3$ (Theorem \ref{POA_unbounded_uniform_m>=3}). When $\kappa$ is odd, the price of stability is $1$ for any $m\geq 2$ (Theorem \ref{ubposuniform}), while the price of anarchy is $3$ for $m=2$ (Theorems \ref{ubpoa} and \ref{lowerbound_poa_uniform_k_odd_two_colors}), and unbounded for $m\geq 3$ (Theorem \ref{POA_unbounded_uniform_m>=3}).
We also design an algorithm (Algorithm \ref{alg:uniformne}) which returns a Nash equilibria which is socially optimal when $\kappa$ is odd and $2$-approximates the social optimal when $\kappa$ is even.

Due to space constraints, some proofs have been removed. All the details will appear in the full version of the paper. 

\noindent{\bf Related Work.}
The classical one-dimensional bin packing problem has been widely studied (see \cite{CGJ96} for a general survey). Bin packing games under the proportional cost function have been introduced in \cite{B06}. The author proved the existence of Nash equilibria by showing that the best-response dynamics converge in finite time. He also established that there is always a Nash equilibrium with minimal number of bins, i.e., the $\pos$ is 1, but that finding such a good equilibrium is NP-hard. Finally, he presented upper and lower bounds on the $\poa$. Nearly tight bounds on the $\poa$ have been later shown in \cite{EK11}. Yu and Zhang \cite{YZ08} have designed a polynomial time algorithm which returns a Nash equilibrium. Bin packing games under the egalitarian cost function were considered in \cite{MDHTYZ13}. They showed tight bounds on the $\poa$ and the $\pos$ and design a polynomial time algorithm for computing a Nash equilibrium. In \cite{DE14}, the authors provided tight bounds on the exact worst-case number of steps needed to reach a Nash equilibrium. Other types of equilibria (like for instance strong equilibria) and other bin packing games were also considered in \cite{AE13,CY11,DE12,E13,EK11,EK15,FFMW11}.

The offline version of the black and white bin packing problem was considered in \cite{BBDEKLT15}. Most of the literature on colorful bin packing is about the online version of the problem. Competitive algorithms for the online colorful bin packing problem were presented in \cite{DE14b}. The special case black and white was considered in \cite{BBDEKT15}, while, the one where all items have size $0$, was considered in \cite{BSV14}. All such results on the online version of the problem were improved in \cite{BDESV}.

Related colorful bin packing problems have been also considered.
For instance, in the bin coloring \cite{KPRS08}, the problem is to pack colored items into bins, such that the maximum number of different colors per bin is minimized. The bin coloring games were considered in \cite{EKLS11}, where players control colored items and each player aims at packing its item into a bin with as few different colors as possible.

To the best of our knowledge, this is the first paper dealing with Nash equilibria in colorful bin packing games.

\section{Model and Preliminaries}
In a {\em colorful bin packing game} $G=(N,C,{\cal B},(s_i)_{i\in N},(c_i)_{i\in N})$ we have a set of $n$ players $N=\{1,\ldots,n\}$, a set of $m\geq 2$ colors $C=\{1,\ldots,m\}$ and a set of $n$ unit capacity bins ${\cal B}=\{B_1\ldots,B_n\}$. Each player $i\in N$ controls an indivisible item, denoted for convenience as $x_i$ (i.e., we denote by $X=\{x_1,\ldots,x_n\}$ the set of items), having size $s_i\in [0,1]$ and color $c_i\in C$ which needs to be packed into one bin in $\cal B$ without exceeding its capacity. Game $G$ has {\em uniform sizes} if 
$s_i=s_j$ for every $i,j\in N$.
The special case in which $G$ has $m=2$ colors is called the {\em black and white bin packing game}; we shall define color $1$ as black, color $2$ as white and denote by $\#B$ and $\#W$ the number of black and white items in $G$, respectively.

A strategy profile is modeled by an $n$-tuple ${\bm\sigma}=(\sigma_1, \ldots, \sigma_n)$ such that, for each $i\in N$, $\sigma_i\in {\cal B}$ is the bin chosen by player $i$.
We denote by $B_j({\bm\sigma})=\{x_i \in X : \sigma_i=B_j\}$ the set of items packed into $B_j$ according to the strategy profile ${\bm\sigma}$. Similarly, we also write 
$\sigma_i({\bm\sigma})=\{x_l \in X : \sigma_l=\sigma_i\}$ to indicate the set of items packed in the same bin as $x_i$ (i.e., the bin chosen by player $i$), according to ${\bm\sigma}$.
%
Given any bin $B_j$, we assume to pack the items in a fixed internal order, going from bottom to top, that is the sequential order in which players have chosen the bin $B_j$ as strategy. 
Namely, for any pair of items $x_i$ and $x_l$ in $B_j({\bm\sigma})$, we say that $x_i$ precedes $x_l$ inside the bin $B_j$, and we write $x_i \prec_{\bm\sigma} x_l$, if player $i$ chose bin $B_j$ before $l$. Formally, given any strategy profile ${\bm\sigma}$, each item $x_i$ occupies a precise position $p_i(\bm\sigma)$ in the sequential order of items in bin $\sigma_i$, counting from bottom to top, computed as $p_i(\bm\sigma)=1 + |\{x_l\in \sigma_i(\bm\sigma): x_l \prec_{\bm\sigma} x_i \}|$.
For convenience, if the strategy profile is clear from the context we may omit the symbol ${\bm\sigma}$ and simply write $p_i$.
We notice that, with such packing, the last player, say $i$, choosing the bin $\sigma_i$, occupies the top position in $\sigma_i$.  

Denoted by $\ell_{B_j}({\bm\sigma})=\sum_{x_i\in B_j({\bm\sigma})}s_i$ the total size of items packed into $B_j({\bm\sigma})$, we always assume that $\ell_{B_j}({\bm\sigma})\leq 1$, so that every strategy profile induces a packing of items in $\cal B$ and vice versa.

We say that an item $x_i$ is {\em misplaced} if there exists an item $x_{l}$ with $c_i=c_{l}$ such that $\sigma_i=\sigma_{l}$ and $|p_i-p_l|=1$, that is, $x_i$ is packed next to an item of the same color. 
A bin is {\em feasible} if it stores no misplaced items. In particular, an empty bin is feasible. A strategy profile is feasible if so are all of its bins. For games with uniform sizes $s_i=s$ for every $i\in N$, we denote by $\kappa = \left\lfloor \frac{1}{s} \right\rfloor$ the maximum number of items that can be packed into any (even non-feasible) bin. We only consider the cases in which $\kappa>1$ as, otherwise, the game is trivial.

We shall denote by $cost_i({\bm\sigma})$ the cost that player $i\in N$ pays in the strategy profile $\bm\sigma$ and each player aims at minimizing it. We consider two different cost functions: the {\em egalitarian cost function} and the {\em proportional cost function}. We have $cost_i({\bm\sigma})=\infty$ under both cost functions when $x_i$ is a misplaced item, while, for non-misplaced ones, we have $cost_i({\bm\sigma})=\frac{1}{|{\sigma_i}({\bm\sigma})|}$ under the egalitarian cost function and $cost_i({\bm\sigma})=\frac{s_i}{\ell_{\sigma_i}({\bm\sigma})}$ under the proportional one. Note that, for games with uniform sizes, the two cost functions coincide. For a fixed strategy profile $\bm\sigma$, we say that a bin is a {\em singleton} bin if it stores only one item. Moreover, when considering the egalitarian (resp. proportional) cost function, we denote by $\overline{B}_{\bm\sigma}$ the bin storing the maximum number of items (resp. the fullest bin) in the packing corresponding to $\bm\sigma$, breaking ties arbitrarily.

A {\em deviation} for a player $i$ in a strategy profile $\bm\sigma$ is the action of changing the selected bin $\sigma_i$ in favor of another bin, say $B_j$, such that $\ell_{B_j}({\bm\sigma})+s_i\leq 1$. 
We shall denote as $({\bm\sigma}_{-i},B_j)$ the strategy profile realized after the deviation. Formally, ${\bm\sigma'}=({\bm\sigma}_{-i},B_j)=(\sigma'_1, ,\ldots, \sigma'_n)$ is defined as follows: $\sigma'_i=B_j$ and $\sigma'_{l}=\sigma_{l}$ for each player $l\neq i$. 

In this paper, we consider deviations of the following form: $x_i$ is removed from $\sigma_i$ and packed on top of $B_j$, consistently with the sequential order of items in a bin. 
%
%

An {\em improving deviation} for a player $i$ in a strategy profile $\bm\sigma$ is a deviation towards a bin $B_j$ such that $cost_i({\bm\sigma}_{-i},B_j)<cost_i({\bm\sigma})$. Fix a feasible strategy profile $\bm\sigma$. Under the egalitarian cost function, player $i$ admits an improving deviation in $\bm\sigma$ if there exists a bin $B_j\in {\cal B}\setminus\{\sigma_i\}$ such that $(i)$ the item on top of $B_j$ has a color different than $c_i$ and $(ii)$ $|{\sigma_i}({\bm\sigma})|\leq |B_j({\bm\sigma})|$. Under the proportional cost function, player $i$ admits an improving deviation in $\bm\sigma$ if there exists a bin $B_j\in {\cal B}\setminus\{\sigma_i\}$ such that $(i)$ the item on top of $B_j$ has a color different than $c_i$ and $(ii)$ $\ell_{\sigma_i}({\bm\sigma}) < \ell_{B_j}({\bm\sigma})+s_i$. Conversely, when a strategy profile $\bm\sigma$ is unfeasible, under both cost functions, a player controlling a misplaced item $x$ always possesses an improving deviation, for instance, by moving $x$ to an empty bin which is always guaranteed to exist as there are $n$ items, $n$ bins and the bin storing $x$ is non-singleton. We note that, as a side-effect of an improving deviation, $({\bm\sigma}_{-i},B_j)$ may be unfeasible even if $\bm\sigma$ is feasible: this happens when $x_i$ separates two items of the same color. We say that an improving deviation is {\em valid} whenever the destination bin is feasible before the deviation.

A strategy profile $\bm\sigma$ is a (pure) Nash equilibrium if $cost_i({\bm\sigma})\leq cost_i({\bm\sigma}_{-i},B_j)$ for each $i\in N$ and $B_j\in {\cal B}$, that is, no  player has an improving deviation in $\bm\sigma$. Let ${\sf NE}(G)$ denote the set of Nash equilibria of game $G$. It is easy to see that any Nash equilibrium is a feasible strategy profile. This implies that $m=1$ would force each item to be packed into a different bin: this justifies our choice of $m\geq 2$. A game $G$ has the {\em finite improvement path property} if it does not admit an infinite sequence of improving deviations. Clearly, if $G$ enjoys the finite improvement path property, it follows that ${\sf NE}(G)\neq\emptyset$.

Given a strategy profile $\bm\sigma$, let $O({\bm\sigma})\subseteq {\cal B}$ be the set of open bins in $\bm\sigma$, where a bin is {\em open} if it stores at least one item. Let ${\sf F}({\bm\sigma})$ be the number of open bins in $\bm\sigma$, i.e.,  ${\sf F}({\bm\sigma}) = |O({\bm\sigma})|$. We shall denote with ${\bm\sigma}^*(G)$ the {\em social optimum}, that is, any strategy profile minimizing function {\sf F}. It is easy to see that any social optimum is a feasible strategy profile.

The {\em price of anarchy} of $G$ is defined as $\poa(G)=\max_{{\bm\sigma}\in{\sf NE}(G)}\frac{{\sf F}({\bm\sigma})}{{\sf F}({\bm\sigma}^*(G))}$, while the {\em price of stability} of $G$ is defined as $\pos(G)=\min_{{\bm\sigma}\in{\sf NE}(G)}\frac{{\sf F}({\bm\sigma})}{{\sf F}({\bm\sigma}^*(G))}$. Given a class of colorful bin packing games $\cal C$, the prices of anarchy and stability of ${\cal C}$ are defined as $\poa({\cal C})=\sup_{G\in {\cal C}}\poa(G)$ and $\pos({\cal C})=\sup_{G\in {\cal C}}\pos(G)$. Let ${\cal G}_m$ denote the set of all colorful bin packing games with $m$ colors and ${\cal U}^{odd}_m$ (resp. ${\cal U}^{even}_m$) denote the set of all colorful bin packing games with $m$ colors and uniform sizes for which $\kappa$ is odd (resp. even). Finally, denote ${\cal U}_m={\cal U}^{even}_m\cup{\cal U}^{odd}_m$.

\section{Existence and Efficiency of Nash Equilibria in General Games}

In this section, we first show that, without any particular restriction on the type of improving deviations performed by the players, even games with uniform sizes and only two colors may not admit the finite improvement path property (Proposition \ref{prop1}). However, if one allows the players to perform only valid improving deviations, then any game possesses the finite improving path property under both cost functions (Theorems \ref{thm1} and \ref{thm2}). These two theorems, together with the fact that in any strategy profile which is not a Nash equilibrium there always exists a valid improving deviation, imply the existence of Nash equilibria for colorful bin packing games under both cost functions.

\begin{proposition}\label{prop1}
There exists a black and white bin packing game with uniform sizes not possessing the finite improvement path property.
\end{proposition}

\begin{IEEEproof}
Let $G$ be a black and white bin packing game with uniform sizes defined by three black items, denoted as $b_1$, $b_2$ and $b_3$, and three white items, denoted as $w_1$, $w_2$ and $w_3$. All items have size $1/4$, so that $\kappa=4$. Figure \ref{fig1} depicts a cyclic sequence of (non-valid) improving deviations which shows the claim.
\begin{figure}[ht]
\centering
\includegraphics[scale=0.35]{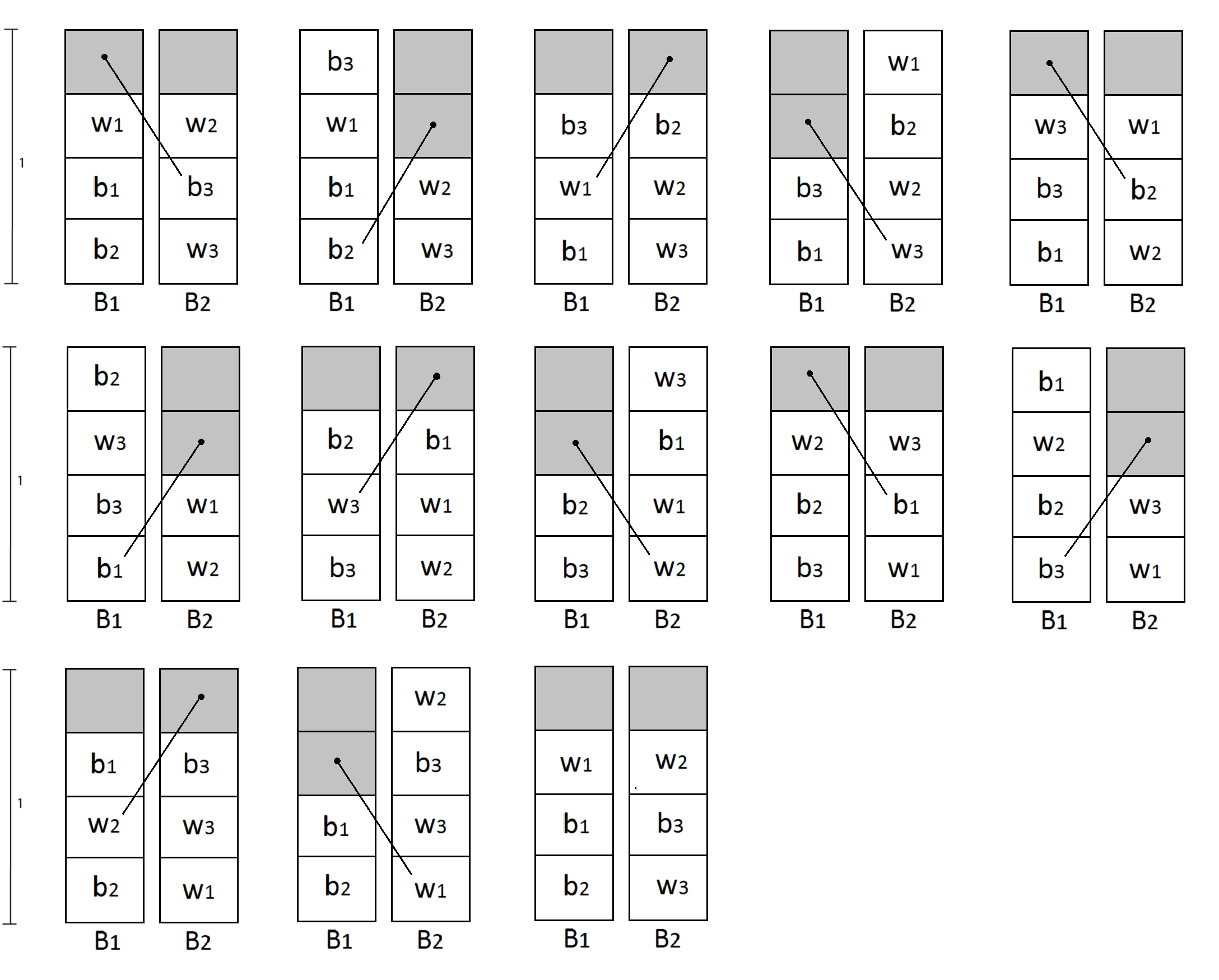}
\caption{A cyclic sequence of (non-valid) improving deviations.}
\label{fig1}
\end{figure}
\end{IEEEproof}

\begin{theorem}\label{thm1}
If players are restricted to perform only valid improving deviations, then each colorful bin packing game under the egalitarian cost function admits the finite improvement path property.
\end{theorem}
\begin{IEEEproof}
To prove the claim, we define a suitable potential function which strictly increases each time a player performs a valid improving deviation.
Given a strategy profile $\bm\sigma$, consider the potential function
$$\Phi({\bm\sigma})=\sum_{B_j\in O({\bm\sigma}):B_j\textrm{ is feasible}}|B_j({\bm\sigma})|^{|B_j({\bm\sigma})|}$$
and assume that a player $i$ performs a valid improving deviation by moving $x_i$ onto bin $B_j\neq\sigma_i$. We distinguish between two cases.

If bin $\sigma_i$ is not feasible
since both $B_j({\bm\sigma})$ and $B_j({\bm\sigma}_{-i},B_j)$ are feasible, we obtain $\Phi({\bm\sigma}_{-i},B_j)-\Phi({\bm\sigma}) \geq (|B_j({\bm\sigma})|+1)^{|B_j({\bm\sigma})|+1}-|B_j({\bm\sigma})|^{|B_j({\bm\sigma})|}>0$ when $B_j({\bm\sigma})$ is open and $\Phi({\bm\sigma}_{-i},B_j)-\Phi({\bm\sigma}) \geq 1^1 -0  > 0 $ otherwise.

If $\sigma_i$ is feasible, under the egalitarian cost function, this implies that $|B_j({\bm\sigma})|\geq |{\sigma_i}({\bm\sigma})|$. Observe that, since $\sigma_i$ is feasible, $cost_i({\bm\sigma})\leq 1$, which implies that it must be $|B_j({\bm\sigma})|\geq 1$. For the ease of notation, set $|{\sigma_i}({\bm\sigma})|=\alpha$ and $|B_j({\bm\sigma})|=\beta$, so that $\beta\geq\alpha$ and $\beta\geq 1$; we get
\begin{eqnarray*}
\Phi({\bm\sigma}_{-i},B_j)-\Phi({\bm\sigma}) & \geq & (\beta+1)^{\beta+1}-\alpha^\alpha-\beta^\beta\\
& = & (\beta+1)(\beta+1)^{\beta}-\alpha^\alpha-\beta^\beta\\
& \geq & 2(\beta+1)^{\beta}-\alpha^\alpha-\beta^\beta\\
& \geq & 2(\beta+1)^{\beta}-2\beta^\beta\\
& > & 0,
\end{eqnarray*}
where the second inequality comes from $\beta\geq 1$ and the third one comes from $\beta\geq\alpha$. Thus, in any case, $\Phi({\bm\sigma}_{-i},B_j)>\Phi({\bm\sigma})$ which, since the number of possible strategy profiles is finite, implies the claim.
\end{IEEEproof}

The next proof uses a different function than the one used above.
\begin{theorem}\label{thm2}
If players are restricted to perform only valid improving deviations, then each colorful bin packing game under the proportional cost function admits the finite improvement path property.
\end{theorem}

\begin{IEEEproof}
To prove the claim, we define a suitable potential function which strictly increases each time a player performs a valid improving deviation. Denote ${\cal P}(N)$ as the power set of $N$ and, given a set $X\in{\cal P}(N)$, denote $S(X)=\sum_{i\in X}s_i$. Define $$\delta=\min_{X,Y\in{\cal P}(N):S(X)\neq S(Y)}\left|S(X)-S(Y)\right|>0$$ and let $\rho>>1$ be a number such that $\rho^\delta>2$. Given a strategy profile $\bm\sigma$, consider the potential function $$\Phi({\bm\sigma})=\sum_{B_j\in O({\bm\sigma}):B_j\textrm{ is feasible}}\rho^{\ell_{B_j({\bm\sigma})}}$$
and assume that a player $i$ performs a valid improving deviation by moving $x_i$ onto bin $B_j\neq\sigma_i$, which implies $s_i>0$.
We distinguish between two cases.

If bin $\sigma_i$ is not feasible,
since both $B_j({\bm\sigma})$ and $B_j({\bm\sigma}_{-i},B_j)$ are feasible, we obtain $\Phi({\bm\sigma}_{-i},B_j)-\Phi({\bm\sigma})\geq \rho^{\ell_{B_j({\bm\sigma})}+s_i}-\rho^{\ell_{B_j({\bm\sigma})}}>0$.

If $\sigma_i$ is feasible,
under the proportional cost function, this implies that $\ell_{B_j}({\bm\sigma})+s_i>\ell_{\sigma_i}({\bm\sigma})$. For the ease of notation, set $\ell_{\sigma_i}({\bm\sigma})=\alpha$ and $\ell_{B_j}({\bm\sigma})=\beta$, so that $\beta+s_i>\alpha$, which, by the definition of $\delta$, implies that $\beta+s_i\geq\max\{\alpha,\beta\}+\delta$; we get
\begin{eqnarray*}
\Phi({\bm\sigma}_{-i},B_j)-\Phi({\bm\sigma}) & \geq & \rho^{\beta+s_i}-\rho^\alpha-\rho^\beta\\
& \geq & \rho^{\max\{\alpha,\beta\}+\delta}-2\rho^{\max\{\alpha,\beta\}}\\
& = & \rho^{\max\{\alpha,\beta\}}\left(\rho^\delta-2\right)\\
& > & 0,
\end{eqnarray*}
where the last inequality comes from the definition of $\rho$. Thus, in any case, $\Phi({\bm\sigma}_{-i},B_j)>\Phi({\bm\sigma})$ which, since the number of possible strategy profiles is finite, implies the claim.
\end{IEEEproof}


In the following we give a tight characterization of the efficiency of Nash equilibria in colorful bin packing games under both cost functions. This is achieved by giving upper bounds on the $\poa$ and matching or asymptotically matching lower bounds on the $\pos$.
For games with at least three colors, Theorems \ref{POS_unbounded_egalitarian_m>=3} and
\ref{POS_unbounded_proportional_m>=3}  show that, under both cost functions, the $\pos$ can be unbounded, thus, in the worst-case, no efficient Nash equilibria are guaranteed to exist.

\begin{theorem}\label{POS_unbounded_egalitarian_m>=3}
Under the egalitarian cost function, $\pos({\cal G}_{m})$ is unbounded for each $m\geq 3$.
\end{theorem}

\begin{IEEEproof}
Fix a value $m\geq 3$. We show the theorem by proving that there exists a game $G_{m}\in{\cal G}_m$ with $n$ players such that
\begin{equation*}
{\sf PoS}(G_{m})\geq\left\{
\begin{array}{ll}
\frac{n}{32} & \textrm{ if }m=3,\\
\frac{n(2m-5)}{18(m-1)} & \textrm{ if }m\in\{4,5,6\},\\
\frac{n(m-3)}{8(m-1)} & \textrm{ if }m\geq 7.\\
\end{array}\right.
\end{equation*}
By taking the limit for $n$ going to $\infty$, the theorem will follow.

We construct game $G_{m}$ as follows. There are:
\begin{itemize}
\item[$\bullet$] $h k$ white items of size $1/k- h \delta$,
\item[$\bullet$] for every color other than white, $\frac{h (k-1)}{m-1}$ items of size $\delta$,
\end{itemize}
where $h$ and $k$ are arbitrary integers such that $h\geq 2$ and $k$ is a multiple of $m$, while $\delta>0$ is an arbitrarily small real value such that $\delta< (h k(k+1))^{-1}$.

Observe that, since a combination of $k$ white items and $k-1$ non-white items can be feasibly packed into the same bin, and there are exactly $h k$ white items and $h (k-1)$ non-white items, there exists a feasible strategy profile using exactly $h$ bins. Moreover, note that $k+1$ white items do not fit into a bin, as their total size is equal to $1+1/k- h (k+1)\delta>1$ by the definition of $\delta$.

We proceed by showing that any Nash equilibrium for $G_{m}$ needs to use at least $k\left(h-1-\frac{h}{m-1}\right)$ bins. To this aim, fix a Nash equilibrium $\bm\sigma$ for $G_{m}$ and consider bin $\overline{B}_{\bm\sigma}$.

If the item on top of $\overline{B}_{\bm\sigma}$ is white, then all non-white items have to be stored in $\overline{B}_{\bm\sigma}$. In fact, since $\overline{B}_{\bm\sigma}$ can contain at most $k$ white items, for an overall occupation of $1-h k \delta$, it follows that $\overline{B}_{\bm\sigma}$ has enough space to accommodate all the $h (k-1)$ non-white items. Thus, since $\overline{B}_{\bm\sigma}$ stores all the non-white items and can store at most $k$ white items, it follows that $\bm\sigma$ needs to use at least $k(h -1)$ singleton bins to feasibly store all white items not stored in $\overline{B}_{\bm\sigma}$, so that ${\sf F}({\bm\sigma})\geq 1+k(h-1)\geq k\left(h -1-\frac{h}{m-1}\right)$.

Hence, assume that the item of top of $\overline{B}_{\bm\sigma}$ has a color $c$ which is not white. By the same arguments exploited above, it follows that $\overline{B}_{\bm\sigma}$ have to contain all non-white items having color different than $c$. Again, since $\overline{B}_{\bm\sigma}$ can contain at most $k$ white items, $\bm\sigma$ needs to pack at least $k(h -1)$ white items outside bin $\overline{B}_{\bm\sigma}$. To do this, at most $\frac{h (k-1)}{m-1}-1$ non-white items (the ones having color $c$) can be used. The only way to pack these items so that $\bm\sigma$ results in a Nash equilibrium is to use a first-fit-like algorithm. Note that, in order to pack $k$ white items in the same bin, at least $k-1$ item of color $c$ are needed. Hence, by using $\frac{h (k-1)}{m-1}-1$ items of color $c$, at most $\frac{h k}{m-1}$ white items can be packed in a non-singleton bin, so that at least $k\left(h-1-\frac{h}{m-1}\right)$ white items need to be stored in a singleton bin.

Thus, we can conclude that ${\sf PoS}(G_{m})\geq\frac{k}{h} \left(h-1-\frac{h}{m-1}\right)$. Set $h=4$ for $m=3$, $h=3$ for $m\in\{4,5,6\}$, and $h=2$ for $m\geq 7$. For sufficiently high values of $k$, by using $k>\frac{n}{2 h}$, the claim follows.
\end{IEEEproof}

\begin{theorem}\label{POS_unbounded_proportional_m>=3}
Under the proportional cost function, $\pos({\cal G}_{m})$ is unbounded for each $m\geq 3$.
\end{theorem}
\begin{IEEEproof}
We prove the claim under the hypothesis of $m = 3$. It is easy to adapt the proof so
as to deal with any number of colors $m \geq 3$.

Consider the following instance with $n$ players, where $n$ is a multiple of $4$. There are $\frac{n}{2}$ items of color $c_1$, $\frac{n}{4}$ of color $c_2$, and $\frac{n}{4}$ of color $c_3$. The sizes are such that, one item of color $c_1$ has size $a$, while all the others $\frac{n}{2} -1$ ones have size $b$. Finally, each item of color $c_2$ or $c_3$ has size $c$. The values of the sizes $a,b,c$ are as follows:
\begin{itemize}
\item $a=1-\frac{2}{n} + \epsilon$,  for any $0<\epsilon<\frac{2}{n}$ (i.e., $a<1$).
\item $b=1-a$
\item $0< c \leq \frac{2}{n}(\frac{2}{n} - \epsilon)$
\end{itemize}

Notice that $\sum_{i=1}^n s_i > 1$, therefore any solution cannot use less than two bins. An optimal solution can be achieved by assigning the item of size $a$ to one bin, and all the other items to a second bin, as depicted in Figure \ref{propPoS}. Notice that, {\em (i)} the total size of items packed in the second bin is at most $a$. Moreover, {\em (ii)} $a + c\frac{n}{2} \leq 1$, i.e., the item of size $a$ and all the items of size $c$ can be packed together in a bin.

We further notice that, the item of size $a$ cannot be packed with any other item of the same color $c_1$, because at least one element of different color is needed between them, and $a+b+c>1$.

Consider any Nash equilibrium where the item of size $a$ is packed in a bin $B$. We now show that at most two items of size $c$, are not packed in $B$. Let us suppose, by contradiction, that there exist at least three items of size $c$ that are not packed in the bin $B$. If among such three items, there exist one of color $c_2$ and another one of color $c_3$, then, by properties {\em (i)} and {\em (ii)}, we get a contradiction with the fact that it is a Nash Equilibrium. Thus, the three items must have the same color, and without loss of generality,  suppose that it is $c_2$. In this case we also get a contradiction. Indeed, it is easy to see that, it is not possible to pack $\frac{n}{2}$ items of color $c_3$ in the bin $B$, by using at most $\frac{n}{2}-3$ items of color $c_2$. We conclude by noticing that, at least $\frac{n}{2} - 5$ items of sizes $b$ must be packed into singleton bins, and this concludes the proof.
\begin{figure}[H]
\centering
\includegraphics[scale=0.5]{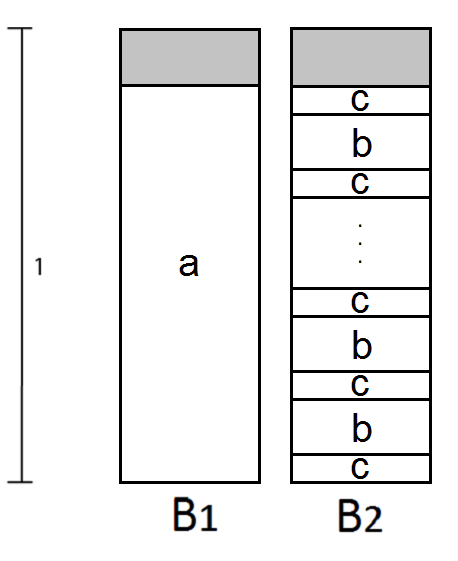}
\caption{An optimal configuration for the instance considered in Theorem \ref{POS_unbounded_proportional_m>=3}.}
\label{propPoS}
\end{figure}
\end{IEEEproof}

We conclude the section by presenting a simple algorithm, namely Algorithm \ref{alg:ne}, for computing a Nash equilibrium in colorful bin packing games under both cost functions. In particular, we shall prove that its running time is polynomial for the egalitarian cost function and pseudo-polynomial for the proportional one for the special case of constant number of colors. Algorithm \ref{alg:ne} is based on the computation of a solution for the following two optimization problems.

\

\noindent{\sf Max Cardinality Colorful Packing}: Given a set of items $X=\{x_1,\ldots,x_n\}$, where, for each $1\leq i\leq n$, $x_i$ has size $s_i\in [0,1]$ and color $c_i$, compute a set of items, of maximum cardinality, which can be packed into a feasible bin without exceeding its capacity.

\

\noindent{\sf Colorful Subset Sum}: Given a set of items $X=\{x_1,\ldots,x_n\}$, where, for each $1\leq i\leq n$, $x_i$ has size $s_i\in [0,1]$ and color $c_i$, compute a set of items, of maximum total size, which can be packed into a feasible bin without exceeding its capacity.

\


\alglanguage{pseudocode}
\begin{algorithm*}[h]
\caption{It takes as input a colorful bin packing game $G$}\label{alg:ne}
\begin{algorithmic}[1]
\State {$X\leftarrow \{x_1,\ldots,x_n\}$}
\While {$(X\neq\emptyset)$}
\If {($G$ is defined under the egalitarian cost function)}
\State {Let $B$ be a solution to {\sf Max Cardinality Colorful Packing}($X$)}
\Else
\State {Let $B$ be a solution to {\sf Colorful Subset Sum}($X$)}
\EndIf
\State {Open a new bin and assign it the set of items $B$}
\State {$X\leftarrow X\setminus B$}
\EndWhile
\State \Return{the strategy profile induced by the set of open bins}
\end{algorithmic}
\end{algorithm*}

Next lemma shows the correctness of the algorithm.

\begin{lemma}\label{compne}
Algorithm \ref{alg:ne} computes a Nash equilibrium for any colorful bin packing game $G$.
\end{lemma}

\begin{IEEEproof}
Consider the strategy profile $\bm\sigma$ returned by Algorithm \ref{alg:ne} and assume, for the sake of contradiction, that it is not a Nash equilibrium. Then, there exist an item $x_i$ packed into a bin $B_j$ and a bin $B_k\neq B_j$ such that $x_i$ can be feasibly packed into $B_k$ and
$|B_j({\bm\sigma})|\leq |B_k({\bm\sigma})|$ under the egalitarian cost function, while $\ell_{B_j}({\bm\sigma})<\ell_{B_k}({\bm\sigma})+s_i$ under the proportional one.
Now, if $B_j$ is opened before $B_k$ by Algorithm \ref{alg:ne}, we have that the set of items packed into $B_k$ together with $x_i$ is a better packing than $B_j$, thus contradicting its optimality. Conversely, if $B_k$ is opened before $B_j$ by Algorithm \ref{alg:ne}, we have that the set of items packed into $B_k$ together with $x_i$ is a better packing than $B_k$, thus contradicting its optimality. Hence, in both cases, we get a contradiction.
\end{IEEEproof}

\noindent We are also able to show that {\sf Max Cardinality Colorful Packing} can be efficiently solved.

\begin{lemma}\label{poly-proc}
{\sf Max Cardinality Colorful Packing} can be solved in polynomial time.
\end{lemma}

\begin{IEEEproof}
Our proof is based on the following two fundamental observations:
{\em (i)} a set $B$ of colored items can be feasibly packed into a bin without exceeding its capacity if and only if the frequency of the dominant color in $B$ is at most $\lceil|B|/2\rceil$, {\em (ii)} if there exists a set $B$ of colored items that can be feasibly packed into a bin without exceeding its capacity whose dominant color $c^*$ has frequency $k^*$, then the set of items computed as follows:
\begin{itemize}
\item[$\bullet$] choose the $k^*$ items of color $c^*$ having minimum size;
\item[$\bullet$] let $Y$ be the set of items obtained by choosing from $X$, for each color other than $c^*$, the $k^*$ items of minimum size (or all items if there are less than $k^*$ items of that color);
\item[$\bullet$] number all items in $Y$ in non-decreasing order of sizes;
\item[$\bullet$] choose the first $|B|-k^*$ items in $Y$;
\end{itemize}
can also be feasibly packed into a bin without exceeding its capacity.

Let $B^*$ be an optimal solution to the problem, $c^*$ be the dominant color in $B^*$ and $k^*$ be the frequency of the items of color $c^*$ in $B^*$. If we knew both $c^*$ and $k^*$, because of observation {\em (ii)}, a solution $B'$ such that $|B'|=|B^*|$ can be constructed in time $O(n\log n)$. Thus, the claim follows by guessing all possible $n^3$ pairs $(c^*,k^*,|B^*|)$, with $k^*\leq\lceil|B^*|/2\rceil$, and then returning the best feasible solution.\footnote{Indeed, it is possible to lower the computational complexity of the algorithm by considering only the $n^2$ pairs $(c^*,k^*)$ by noting that the algorithm used within observation {\em (ii)} can be easily adapted to deal with the case in which $|B|$ is not known.}
\end{IEEEproof}

For what concerns {\sf Colorful Subset Sum}, the problem can be solved in pseudo-polynomial time as long as the number of colors is constant.

\begin{lemma}\label{presudopoly-proc}
{\sf Colorful Subset Sum} can be solved in pseudo-polynomial time as long as the number of colors is constant.
\end{lemma}

\begin{IEEEproof}
Assume to know the exact number of items of each color $c^*_1,\ldots,c^*_m$ that belong to an optimal solution. Note that {\sf Colorful Subset Sum} becomes a variant of a {\sf Knapsack} problem in which item are partitioned into $m$ sets according to their colors and one wants to maximize the sum of the volumes of the items packed in the knapsack without exceeding its capacity by choosing exactly $c^*_i$ items from each set. By suitably extending the dynamic programming algorithm for the {\sf knapsack} problem and guessing the $O(n^m)$ possible tuples $c^*_1,\ldots,c^*_m$ such that $c^*=\max\{c^*_1,\ldots,c^*_m\}\leq\left\lceil\frac{\sum_{i}c^*_i-c^*}{2}\right\rceil$, the claim follows.
\end{IEEEproof}

As a consequence of Lemmas \ref{compne}, \ref{poly-proc} and \ref{presudopoly-proc}, we obtain the following result.

\begin{theorem}\label{thm:computing_equilibria_both_cost_function}
A Nash equilibrium for colorful bin packing games can be computed in polynomial time under the egalitarian cost function and in pseudo-polynomial time for a constant number of colors under the proportional one.
\end{theorem}

\section{Efficiency of Nash Equilibria in Black and White Games}

For black and white bin packing games, things get much more interesting, as we show an upper bound of $3$ on the $\poa$ and a corresponding lower bound on the $\pos$. To address this particular case, given a black and white bin packing game $G$, we make use of the following additional notation. Given a strategy profile $\bm\sigma$, we denote by $S_b({\bm\sigma})$ the set of singleton bins storing a black item, by $S_w({\bm\sigma})$ the set of singleton bins storing a white item, by $M_b({\bm\sigma})$ the set of non-singleton bins having a black item on top, and by $M_w({\bm\sigma})$ the set of non-singleton bins having a white item on top.

The following lemma relates the set of open bins of a feasible strategy profile with that of a social optimum.

\begin{lemma}\label{lemmak}
Fix a feasible strategy profile $\bm\sigma$ and a social optimum ${\bm\sigma}^*$ for a black and white bin packing game $G$. Then, $|S_b({\bm\sigma})|-|S_w({\bm\sigma})|-|M_w({\bm\sigma})|\leq {\sf F}({\bm\sigma}^*)$.
\end{lemma}
\begin{IEEEproof}
First, we observe that, since in any open bin of ${\bm\sigma}^*$ the absolute value of the difference between the number of black and white items is at most $1$, we have
\begin{equation}\label{eq1}
\#B \leq \#W + {\sf F}({\bm\sigma}^*).
\end{equation}
Now, let us denote with $z_b$ the number of black items packed into bins belonging to $M_b({\bm\sigma})\cup M_w({\bm\sigma})$ and with $z_w$ the number of white items packed into bins belonging to $M_b({\bm\sigma})\cup M_w({\bm\sigma})$. We have $z_b=\#B-|S_b({\bm\sigma})|$ and $z_w=\#W-|S_w({\bm\sigma})|$. Moreover, the number of black items packed into bins belonging to $M_b({\bm\sigma})\cup M_w({\bm\sigma})$ is at least the number of white items packed into bins belonging to $M_b({\bm\sigma})\cup M_w({\bm\sigma})$ minus $|M_w({\bm\sigma})|$, so that, by putting all together, we obtain
\begin{equation}\label{eq2}
\#B \geq \#W + \vert S_b({\bm\sigma}) \vert - \vert S_w({\bm\sigma}) \vert - \vert M_w({\bm\sigma}) \vert.
\end{equation}
By combining inequalities (\ref{eq1}) and (\ref{eq2}), the claim follows.
\end{IEEEproof}

The following theorem gives an upper bound on the $\poa$ of black and white bin packing games under both cost functions.

\begin{theorem}\label{ubpoa}
Under both cost functions, ${\sf PoA}({\cal G}_2)\leq 3$.
\end{theorem}
\begin{IEEEproof}
Given a black and white bin packing game $G$ under a certain cost function, fix a Nash equilibrium $\bm\sigma$ and a social optimum ${\bm\sigma}^*$. Let $S=\sum_{i\in N}s_i$ be the sum of the sizes of all the items. Notice that ${\sf F}({\bm\sigma}^*)\geq \lceil S\rceil$. Assume without loss of generality that $|S_b({\bm\sigma})| \geq |S_w({\bm\sigma})|$ (if this is not the case, we simply swap the two colors).

Let $P$ be the set of pairs of bins constructed as follows: each bin in $S_w({\bm\sigma})$ is paired with a bin in $S_b({\bm\sigma})$, each remaining bin in $S_b({\bm\sigma})$ is paired with a bin in $M_w({\bm\sigma})$, finally, all the remaining bins in $M_w({\bm\sigma})$ and all the bins in $M_b({\bm\sigma})$ are joined into pairs until possible. It is easy to check that, for each created pair of bins $(B_j,B_k)$, it must be
\begin{equation}\label{over}
\ell_{B_j}({\bm\sigma})+\ell_{B_k}({\bm\sigma})>1
\end{equation}
under both cost functions, otherwise the hypothesis that $\bm\sigma$ is a Nash equilibrium would be contradicted. Moreover, by (\ref{over}), it follows that $S>|P|$ which implies that ${\sf F}({\bm\sigma}^*)\geq\lceil S\rceil\geq |P|+1$.

\noindent Now two cases may occur:
\begin{itemize}
\item[$\bullet$] no bin in $S_b({\bm\sigma})$ is left unmatched by $P$, which implies that $|S_b({\bm\sigma})|+|S_w({\bm\sigma})|+|M_b({\bm\sigma})|+|M_w({\bm\sigma})|\leq 2|P|+1$, as at most one bin from the set $M_w({\bm\sigma})\cup M_b({\bm\sigma})$ may remain unmatched. Thus, we obtain
\begin{eqnarray*}
{\sf F}({\bm\sigma}) & = & |S_b({\bm\sigma})|+|S_w({\bm\sigma})|+|M_b({\bm\sigma})|+|M_w({\bm\sigma})|\\
& \leq & 2|P|+1\\
& < & 2{\sf F}({\bm\sigma}^*);
\end{eqnarray*}
\item[$\bullet$] at least one bin in $S_b({\bm\sigma})$ is unmatched by $P$, which implies that
$|S_b({\bm\sigma})|+|S_w({\bm\sigma})|+|M_b({\bm\sigma})|+|M_w({\bm\sigma})|\leq 2|P|+1+|S_b({\bm\sigma})|-|S_w({\bm\sigma})|-|M_w({\bm\sigma})|$. Thus, we obtain
\begin{eqnarray*}
{\sf F}({\bm\sigma}) & = & |S_b({\bm\sigma})|+|S_w({\bm\sigma})|+|M_b({\bm\sigma})|+|M_w({\bm\sigma})|\\
& \leq & 2|P|+1+|S_b({\bm\sigma})|-|S_w({\bm\sigma})|-|M_w({\bm\sigma})|\\
& \leq & 2|P|+1+{\sf F}({\bm\sigma}^*)\\
& < & 2{\sf F}({\bm\sigma}^*)+{\sf F}({\bm\sigma}^*)\\
& = & 3{\sf F}({\bm\sigma}^*),
\end{eqnarray*}
where the second inequality comes from Lemma \ref{lemmak}.
\end{itemize}
\end{IEEEproof}

In the next two theorems, we show a matching lower bound on the $\pos$ of black and white bin packing games under both cost functions.

\begin{theorem}\label{thm:lowerbound_pos_twocolors_egalitarian}
Under the egalitarian cost function, $\pos({\cal G}_2)\geq 3$.
\end{theorem}
\begin{IEEEproof}
We prove the theorem by showing that, for any $\epsilon>0$, there exists a black and white bin packing game $G_{\epsilon}\in{\cal G}_2$ such that $\pos(G_{\epsilon})\geq 3-\epsilon$. $G_{\epsilon}$ is defined by the following set of items:
\begin{itemize}
\item[$\bullet$] $2k$ white items having size $\frac{1}{k}-2\delta$, denoted as items of type (1);
\item[$\bullet$] $k/2$ black items having size $1$, denoted as items of type (2);
\item[$\bullet$] $2k$ black items having size $\delta$, denoted as items of type (3);
\item[$\bullet$] $k$ white items having size $0$, denoted as items of type (4);
\end{itemize}
where $k$ is an even integer such that $k\geq\max\left\{\frac{10-4\epsilon}{\epsilon},19\right\}$ and $\delta>0$ is arbitrarily small.

Denote with $\bm\sigma$ the strategy profile such that $O({\bm\sigma})=(B_1,\ldots,B_h)$ with $h=\frac{3k}{2}+1$ and such that
\begin{itemize}
\item[$\bullet$] bin $B_1$ contains $k$ items of type (1), $k$ items of type (4) and $2k$ items of type (3),
\item[$\bullet$] each bin from $B_2$ to $B_{k+1}$, for a total of $k$ bins, contains one item of type (1),
\item[$\bullet$] each bin from $B_{k+2}$ to $B_h$, for a total of $k/2$ bins, contains one item of type (2).
\end{itemize}
In the definition of $\bm\sigma$, we avoid considering the order in which the items are packed within each bin as it is irrelevant to our purposes. We only stress the fact that there exists a proper ordering of the items which makes $\bm\sigma$ a feasible strategy profile.

Now, denote with ${\bm\sigma}^*$ the strategy profile such that $O({\bm\sigma}^*)=(B^*_1,\ldots,B^*_{h^*})$ with $h^*=\frac{k}{2}+2$ and such that
\begin{itemize}
\item[$\bullet$] bins $B^*_1$ and $B^*_2$ both contain $k$ items of type (1) and $k$ items of type (3),
\item[$\bullet$] each bin from $B^*_3$ to $B^*_{h^*}$, for a total of $k/2$ bins, contains one item of type (2) and 2 items of type (4).
\end{itemize}
Also in this case, we avoid considering the order in which the items are packed within each bin and stress the fact that there exists a proper ordering of the items which makes ${\bm\sigma}^*$ a feasible strategy profile.

In order to show the claimed lower bound on $\pos(G_\epsilon)$, we proceed by proving that the packing of items corresponding to any Nash equilibrium for $G_{\epsilon}$ coincides with the one corresponding to $O(\bm\sigma)$. This is achieved by exploiting a sequence of results.

First of all, we observe the following basic fact.

\begin{fact}\label{f1}
In any Nash equilibrium for $G_{\epsilon}$, any bin containing an item of type (2) can store at most $3$ items.
\end{fact}

We continue by proving some basic properties possessed by bin $\overline{B}_{\bm\tau}$, for any Nash equilibrium $\bm\tau$ for $G_{\epsilon}$.

\begin{lemma}\label{l0}
Fix a Nash equilibrium $\bm\tau$ for $G_{\epsilon}$. Then, $\overline{B}_{\bm\tau}$ stores at least 4 items.
\end{lemma}
\begin{IEEEproof}
Assume, for the sake of contradiction, that $\overline{B}_{\bm\tau}$ stores at most 3 items. Since the total number of items is $\frac{11k}{2}$, it follows that $\bm\tau$ is made up of at least $\left\lceil\frac{11k}{6}\right\rceil$ bins. This implies that
${\sf PoA}(G_{\epsilon})\geq\frac{{\sf F}({\bm\tau})}{{\sf F}({\bm\sigma}^*)}\geq\frac{\left\lceil\frac{11k}{6}\right\rceil}{\frac{k}{2}+2}\geq\frac{\frac{11k}{6}}{\frac{k}{2}+2}=\frac{11}{3}-\frac{44}{3(k+4)}= 3+\frac 2 3-\frac{44}{3(k+4)}>3$. 

Since $k>18$: a contradiction to Theorem \ref{ubpoa}.
\end{IEEEproof}

As a consequence of Fact \ref{f1} and Lemma \ref{l0}, we get the following corollary.

\begin{corollary}\label{c1}
Fix a Nash equilibrium $\bm\tau$ for $G_{\epsilon}$. No item of type (2) is packed into $\overline{B}_{\bm\tau}$.
\end{corollary}

\begin{lemma}\label{l1}
Fix a Nash equilibrium $\bm\tau$ for $G_{\epsilon}$. All items of type (4) are packed into $\overline{B}_{\bm\tau}$.
\end{lemma}
\begin{IEEEproof}
Assume, for the sake of contradiction, that there exists an item of type (4), say $x_i$, which is not packed into $\overline{B}_{\bm\tau}$. Since $\bm\tau$ is a Nash equilibrium, the item on top of $\overline{B}_{\bm\tau}$ must be white, otherwise player $i$ would lower her cost by migrating to $\overline{B}_{\bm\tau}$. By Corollary \ref{c1}, $\overline{B}_{\bm\tau}$ can only store items of type (1), (3) and (4). The load coming from items of type (1) packed into $\overline{B}_{\bm\tau}$ can be at most $1-2k\delta$ (since at most $k$ items of type (1) can be packed into the same bin), so that bin $\overline{B}_{\bm\tau}$ can potentially store all items of type (3). This implies that all of these items must indeed be packed into $\overline{B}_{\bm\tau}$, otherwise the player owning any of the leftover ones would lower her cost by migrating to $\overline{B}_{\bm\tau}$. Hence, we can conclude that $\overline{B}_{\bm\tau}$ stores $2k$ black items and the item on top of $\overline{B}_{\bm\tau}$ is white. This implies that $\overline{B}_{\bm\tau}$ has to store at least $2k$ white items. Now, since at most $k$ items of type (1) can be packed into the same bin, $\overline{B}_{\bm\tau}$ needs to store all items of type (4).
\end{IEEEproof}

\begin{lemma}\label{l2}
Fix a Nash equilibrium $\bm\tau$ for $G_{\epsilon}$. All items of type (3) are packed into $\overline{B}_{\bm\tau}$.
\end{lemma}
\begin{IEEEproof}
Assume, for the sake of contradiction, that there exists an item of type (3), say $x_i$, which is not packed into $\overline{B}_{\bm\tau}$. Again, by Corollary \ref{c1}, $\overline{B}_{\bm\tau}$ can only store items of type (1), (3) and (4) and the load coming from items of type (1) packed into $\overline{B}_{\bm\tau}$ can be at most $1-2k\delta$, so that bin $\overline{B}_{\bm\tau}$ can potentially store all items of type (3). Hence, since $\bm\tau$ is a Nash equilibrium, the item on top of $\overline{B}_{\bm\tau}$ must be black, otherwise player $i$ would lower her cost by migrating to $\overline{B}_{\bm\tau}$. This implies that the total load of $\overline{B}_{\bm\tau}$ has to exceed $1-1/k+2\delta$, otherwise any player owning an item of type (1) would lower her cost by migrating to $\overline{B}_{\bm\tau}$. Given that no item of type (2) can be stored in $\overline{B}_{\bm\tau}$, we have that, in order to achieve a load of more than $1-1/k+2\delta$, $\overline{B}_{\bm\tau}$ has to store exactly $k$ items of type (1). Moreover, by Lemma \ref{l1}, all items of type (4) are packed in $\overline{B}_{\bm\tau}$. Hence, we can conclude that $\overline{B}_{\bm\tau}$ stores exactly $2k$ white items and the item on top of $\overline{B}_{\bm\tau}$ is black. This implies that $\overline{B}_{\bm\tau}$ has to store at least $2k$ black items none of which belonging to type (2), that is, $\overline{B}_{\bm\tau}$ has to store all items of type (3).
\end{IEEEproof}

We can finally prove that all Nash equilibria for $G_{\epsilon}$ correspond to the same packing of items.

\begin{lemma}\label{l3}
Fix a Nash equilibrium $\bm\tau$ for $G_{\epsilon}$. Then, $O({\bm\tau})$ and $O({\bm\sigma})$ are equal up to a renumbering of the bins.
\end{lemma}
\begin{IEEEproof}
Fix a Nash equilibrium $\bm\tau$ for $G_{\epsilon}$. By Lemmas \ref{l1} and \ref{l2}, it follows that $\overline{B}_{\bm\tau}$ stores all items of type (3) and (4), that is, $2k$ black items and $k$ white items. Hence, in order to obtain a feasible strategy profile, $\overline{B}_{\bm\tau}$ needs to store at least $k-1$ items of type (1). Moreover, $\overline{B}_{\bm\tau}$ cannot store more than $k$ items of type (1). If $\overline{B}_{\bm\tau}$ stores $k-1$ items of type (1), the player owning any of the leftover items of type (1) would lower her cost by migrating to $\overline{B}_{\bm\tau}$. This implies that $\overline{B}_{\bm\tau}$ needs to store exactly $k$ items of type (1). Since the remaining $k$ items of type (1) and all items of type (2) can only be packed into different bins, it follows that there exists a suitable renumbering of the bins in $\bm\tau$ which gives $O({\bm\tau})=O({\bm\sigma})$.
\end{IEEEproof}

We can conclude our proof by lower bounding the price of stability of $G_{\epsilon}$. Because of Lemma \ref{l3}, we get
$${\pos}(G_{\epsilon})\geq\frac{{\sf F}({\bm\sigma})}{{\sf F}({\bm\sigma}^*)}=\frac{\frac{3k}{2}+1}{\frac{k}{2}+2}=3-\frac{10}{k+4}\geq 3-\epsilon,$$ since $k\geq\frac{10-4\epsilon}{\epsilon}$ implies that $\epsilon\geq\frac{10}{k+4}$.
\end{IEEEproof}

\begin{theorem}\label{thm:lowerbound_pos_twocolors_proportional}
Under the proportional cost function, $\pos({\cal G}_2)\geq 3$.
\end{theorem}

\begin{IEEEproof}
We prove the theorem by showing that, for any $\epsilon>0$, there exists a black and white bin packing game $G_{\epsilon}\in{\cal G}_2$ such that $\pos(G_{\epsilon})\geq 3-\epsilon$. $G_{\epsilon}$ is defined by the following set of items:
\begin{itemize}
\item[$\bullet$] $2k$ white items having size $\frac{1}{k}-3\delta$, denoted as items of type (1);
\item[$\bullet$] $k/2$ black items having size $1-5k\delta$, denoted as items of type (2);
\item[$\bullet$] $2k$ black items having size $\delta$, denoted as items of type (3);
\item[$\bullet$] $k$ white items having size $\delta$, denoted as items of type (4);
\end{itemize}
where $k$ is an even integer such that $k\geq\max\left\{\frac{10-4\epsilon}{\epsilon},2\right\}$ and $\delta>0$ is an arbitrarily small number satisfying $\delta<(k(5k+3))^{-1}$.

Denote with $\bm\sigma$ the strategy profile such that $O({\bm\sigma})=(B_1,\ldots,B_h)$ with $h=\frac{3k}{2}+1$ and such that
\begin{itemize}
\item[$\bullet$] bin $B_1$ contains $k$ items of type (1), $k$ items of type (4) and $2k$ items of type (3),
\item[$\bullet$] each bin from $B_2$ to $B_{k+1}$, for a total of $k$ bins, contains one item of type (1),
\item[$\bullet$] each bin from $B_{k+2}$ to $B_h$, for a total of $k/2$ bins, contains one item of type (2).
\end{itemize}
In the definition of $\bm\sigma$, we avoid considering the order in which the items are packed within each bin as it is irrelevant to our purposes. We only stress the fact that there exists a proper ordering of the items which makes $\bm\sigma$ a feasible strategy profile.

Now, denote with ${\bm\sigma}^*$ the strategy profile such that $O({\bm\sigma}^*)=(B^*_1,\ldots,B^*_{h^*})$ with $h^*=\frac{k}{2}+2$ and such that
\begin{itemize}
\item[$\bullet$] bins $B^*_1$ and $B^*_2$ both contain $k$ items of type (1) and $k$ items of type (3),
\item[$\bullet$] each bin from $B^*_3$ to $B^*_{h^*}$, for a total of $k/2$ bins, contains one item of type (2) and 2 items of type (4).
\end{itemize}
Also in this case, we avoid considering the order in which the items are packed within each bin and stress the fact that there exists a proper ordering of the items which makes ${\bm\sigma}^*$ a feasible strategy profile.

In order to show the claimed lower bound on $\pos(G_\epsilon)$, we proceed by proving that the packing of items corresponding to any Nash equilibrium for $G_{\epsilon}$ coincides with the one corresponding to $O(\bm\sigma)$. This is achieved by exploiting a sequence of results.

\begin{fact}\label{fact1}
No bin can store more than $k$ items of type (1).
\end{fact}
\begin{IEEEproof}
$k+1$ items of type (1) require a total space of $(k+1)(1/k-3\delta)=1+1/k-3(k+1)\delta>1$ since $\delta<(k(5k+3))^{-1}<(3k(k+1))^{-1}$.
\end{IEEEproof}
\begin{fact}\label{fact1bis}
No bin can store more than $1$ item of type (2).
\end{fact}
\begin{IEEEproof}
Two items of type (2) require a total space of $2-10k\delta>1$ since $\delta<(k(5k+3))^{-1}<(10k)^{-1}$.
\end{IEEEproof}
\begin{fact}\label{fact2}
No bin can simultaneously store items of types (1) and (2).
\end{fact}
\begin{IEEEproof}
One item of type (1) and one item of type (2) require a total space of $1/k-3\delta+1-5k\delta=1+1/k-(5k+3)\delta>1$ since $\delta<(k(5k+3))^{-1}$.
\end{IEEEproof}
\begin{fact}\label{fact2bis}
For any strategy profile $\bm\tau$, $\overline{B}_{\bm\tau}$ either contains exactly one item of type (2) and no items of type (1) or exactly $k$ items of type (1) and no items of type (2).
\end{fact}
\begin{IEEEproof}
Clearly, we have
\begin{equation}\label{maxocc}
\ell_{\overline{B}_{\bm\tau}}({\bm\tau})\geq 1-5k\delta
\end{equation}
(a bin containing an items of type (2) has at least this occupation) which, given that $\delta<(k(5k+3))^{-1}<(8k)^{-1}$, $\overline{B}_{\bm\tau}$ cannot contain only items of types (3) and (4). If $\overline{B}_{\bm\tau}$ contains items of type (2), then, by Facts \ref{fact1bis} and \ref{fact2}, it has to contain exactly one item of this type. On the contrary, if $\overline{B}_{\bm\tau}$ contains items of type (1), then by Facts \ref{fact1} and \ref{fact2}, it can only contain $k$ items of this time, but no more than $k$ of them. If $\overline{B}_{\bm\tau}$ contains at most $k-1$ items of type (1), then, as $\overline{B}_{\bm\tau}$ can additionally contain all items of types (3) and (4), we have $\ell_{\overline{B}_{\bm\tau}}({\bm\tau})\leq (k-1)(1/k-3\delta)+3k\delta=1-1/k+3\delta<1-5k\delta$ as $\delta<(k(5k+3))^{-1}$, thus contradicting inequality (\ref{maxocc}).
\end{IEEEproof}
\begin{fact}\label{fact3}
For any strategy profile $\bm\tau$, $\overline{B}_{\bm\tau}$ has enough unused space to store all items of types (3) and (4) which are not packed into $\overline{B}_{\bm\tau}$.
\end{fact}
\begin{IEEEproof}
Fix a strategy profile $\bm\tau$. By Fact \ref{fact2bis}, we need to distinguish between two cases only. If $\overline{B}_{\bm\tau}$ contains an item of type (2) and no items of type (1), since all items of types (3) and (4) can be packed into the leftover space of $5k\delta$, the claim follows. If $\overline{B}_{\bm\tau}$ contains $k$ item of type (1) and no items of type (2), since all items of types (3) and (4) can be packed into the leftover space of $3k\delta$, the claim follows.
\end{IEEEproof}

We continue by showing a fundamental structural property.

\begin{lemma}\label{cisiamo}
Fix a Nash equilibrium $\bm\tau$ for $G_{\epsilon}$. Then, $\overline{B}_{\bm\tau}$ stores all items of type (4).
\end{lemma}
\begin{IEEEproof}
Fix a Nash equilibrium $\bm\tau$ for $G_{\epsilon}$ and assume, by way of contradiction, that at least one item $x_j$ of type (4) is not packed into $\overline{B}_{\bm\tau}$. By Fact \ref{fact3}, it follows that the item on top of $\overline{B}_{\bm\tau}$ is white, otherwise player $j$ would lower her cost by migrating to $\overline{B}_{\bm\tau}$. Again, by Fact \ref{fact3}, this implies that all items of type (3) are packed into $\overline{B}_{\bm\tau}$, otherwise the player owing a leftover item would lower her cost by migrating to $\overline{B}_{\bm\tau}$. Thus, we have that $\overline{B}_{\bm\tau}$ stores at least $2k$ black items. As, by hypothesis $\overline{B}_{\bm\tau}$ does not store all items of type (4), by Fact \ref{fact2bis}, the number of white items packed into $\overline{B}_{\bm\tau}$ is at most $2k-1$. Thus, we get a contradiction as it is not possible to feasible pack $2k$ black items and $2k-1$ white items in such a way that the item of top of the bin is white.
\end{IEEEproof}

We can now prove that all Nash equilibria for $G_{\epsilon}$ correspond to the same packing of items.

\begin{lemma}\label{dai}
Fix a Nash equilibrium $\bm\tau$ for $G_{\epsilon}$. Then, $O({\bm\tau})$ and $O({\bm\sigma})$ are equal up to a renumbering of the bins.
\end{lemma}
\begin{IEEEproof}
Fix a Nash equilibrium $\bm\tau$ for $G_{\epsilon}$. We show the claim by proving that $\overline{B}_{\bm\tau}$ contains $k$ items of type (1) and all items of types (3) and (4). By Fact \ref{fact2bis}, we have to distinguish between two cases only.

Assume first that $\overline{B}_{\bm\tau}$ contains an item of type (2) and no items of type (1). As $\overline{B}_{\bm\tau}$ can feasibly store all items of type (4) and no more than $k$ items of type (3), we get $\ell_{\overline{B}_{\bm\tau}}({\bm\tau})\leq 1-5k\delta+2k\delta=1-3k\delta$. Now, let $\widetilde{B}_{\bm\tau}$ be the bin containing the maximum number of items of type (1) in $\bm\tau$. We claim that $\widetilde{B}_{\bm\tau}$ contains $k$ items of type (1). Assume, by way of contradiction, that $\widetilde{B}_{\bm\tau}$ contains at most $k-1$ items of type (1). If the item on top of $\widetilde{B}_{\bm\tau}$ is black, we get a contradiction as the player owning an item of type (1) not packed into $\widetilde{B}_{\bm\tau}$ would lower her cost by migrating to $\widetilde{B}_{\bm\tau}$. If the item on top of $\widetilde{B}_{\bm\tau}$ is white, it follows that $\widetilde{B}_{\bm\tau}$ can store at most $k-1$ items of type (3). Given that $\overline{B}_{\bm\tau}$ contains at most $k$ items of type (3), it follows that there is an item of this type, say $x_j$ which is packed neither into $\overline{B}_{\bm\tau}$ nor into $\widetilde{B}_{\bm\tau}$. Hence, we get a contradiction as player $j$ would lower her cost by migrating to $\widetilde{B}_{\bm\tau}$. Thus, $\widetilde{B}_{\bm\tau}$ contains $k$ items of type (1). By the same argument exploited above, it also follows that $\widetilde{B}_{\bm\tau}$ also contains $k$ items of type (3). Hence we get $\ell_{\widetilde{B}_{\bm\tau}}({\bm\tau})=k(1/k-3\delta)+k\delta=1-2\delta>\ell_{\overline{B}_{\bm\tau}}({\bm\tau})$ thus contradicting the fact that $\overline{B}_{\bm\tau}$ contains an item of type (2) and no items of type (1). So, we conclude that this case cannot occur.

Now assume that $\overline{B}_{\bm\tau}$ contains $k$ items of type $k$ and no items of type (2). By Lemma \ref{cisiamo}, $\overline{B}_{\bm\tau}$ also contains all $k$ items of type (4), so that it contains $2k$ white items. Moreover, if $\overline{B}_{\bm\tau}$ does not contain all items of type (3), by Fact \ref{fact3}, the item on top of $\overline{B}_{\bm\tau}$ must be black, otherwise the player owing any leftover item would lower her cost my migrating to $\overline{B}_{\bm\tau}$. But this raises a contradiction, since it is not possible to feasible pack $2k$ white items and $2k-1$ black one in such a way that the item on top of the bin is black. Hence, we have that $\overline{B}_{\bm\tau}$ contains all items of type (3) which shows the claim.
\end{IEEEproof}
We can conclude our proof by lower bounding the price of stability of $G_{\epsilon}$. Because of Lemma \ref{dai}, we get
$${\pos}(G_{\epsilon})\geq\frac{{\sf F}({\bm\sigma})}{{\sf F}({\bm\sigma}^*)}=\frac{\frac{3k}{2}+1}{\frac{k}{2}+2}=3-\frac{10}{k+4}\geq 3-\epsilon,$$ since $k\geq\frac{10-4\epsilon}{\epsilon}$ implies that $\epsilon\geq\frac{10}{k+4}$.
\end{IEEEproof}

\section{Efficiency of Nash Equilibria in Games with Uniform Sizes}
In this section, we provide a complete picture of the efficiency of Nash equilibria for games with uniform sizes. We remind the reader that, in this setting, the egalitarian and proportional cost functions are equivalent. For the sake of simplicity, we say that a bin is \textit{full} if it contains $\kappa$ items.

First, we give a lower bound of $2$ on the $\pos$ for games with any number of colors under the hypothesis that $\kappa$ is an even number.

\begin{theorem}\label{lbposuniform}
For each $m\geq 2$, $\pos({\cal U}^{even}_m) \geq 2$.
\end{theorem}

\begin{IEEEproof}
We prove the claim under the hypothesis of $m=2$. It is easy to adapt the proof so as to deal with any number of colors $m\geq 2$. In particular, we show that, for any fixed $\epsilon>0$, there exists a game $G_{\epsilon}\in{\cal U}^{even}_2$ such that $\pos(G_\epsilon)\geq 2-\epsilon$.

Game $G_{\epsilon}$ is defined as follows: there are $n=k(k+1)/2$ items, of which $k^2/4+k/2$ are white and the remaining $k^2/4$ are black, where $k$ is an even number such that $k\geq 2(2-\epsilon)/\epsilon$. The size of each item is set in such a way that $\kappa = k$.

Let ${\bm\sigma}^*$ be the strategy profile such that $O({\bm\sigma}^*)=(B_1,\ldots,B_{k/2+1})$ where
\begin{itemize}
\item[$\bullet$] each of the first $k/2$ bins contains $k/2$ white items and $k/2-1$ black items;
\item[$\bullet$] bin $B_{k/2+1}$ contains $k/2$ white items and $k/2$ black items.
\end{itemize}

We continue by proving that each Nash equilibrium for $G_{\epsilon}$ uses $k$ open bins. Towards this end, assume, by way of contradiction, that there exists a Nash equilibrium with less than $k/2$ full bins. As $k$ is even, each full bin stores exactly $k/2$ white items and $k/2$ black items, so that the set of full bins can store at most $\frac{k^2}{4}-\frac k 2$ items of the same color. It follows that the set of non-full bins have to store at least $k$ white items and at least $k/2$ black ones. Call these items {\em leftover} items. Let $\widetilde{B}$ be the non-full bin with the highest number of items. Clearly, it contains at most $k-1$ items and we consider two cases: if $\widetilde{B}$ has a black item on top, as $k-1$ is odd, $\widetilde{B}$ contains at most $\frac{k}{2}-1$ white items. A player owning a leftover white item not packed into $\widetilde{B}$ (there are $k/2+1$ such items) has an improving deviation to $\widetilde{B}$ which raises a contradiction. If $\widetilde{B}$ has a white item on top, as $k-1$ is odd, $\widetilde{B}$ contains at most $\frac{k}{2}-1$ black items. A player owning the only leftover black item not packed into $\widetilde{B}$ has an improving deviation to $\widetilde{B}$ which, again, raises a contradiction. Thus, each Nash equilibrium for $G_{\epsilon}$ uses at least $k/2$ full bins.

Now note that there are not enough black items in $G_{\epsilon}$ to create $k/2+1$ feasible full bins. Hence, since every Nash equilibrium is feasible, it follows that each Nash equilibrium $\bm\sigma$ has exactly $k/2$ full bins, where a total number of $k^2/4$ white items and $k^2/4$ black ones are packed. The remaining $k/2$ white items can be feasibly packed only into singleton bins, thus yielding ${\sf F}({\bm\sigma})=k$. By the definition of $k$, we get $\pos(G_\epsilon)\geq\frac{k}{k/2+1}\geq 2-\epsilon$.
\end{IEEEproof}

We show that, unlike the case of general games considered in the previous section, under the hypothesis of uniform sizes, efficient Nash equilibria are always guaranteed to exist for any number of colors. In particular, we design an algorithm which, given a colorful bin packing game $G$ with uniform sizes, returns a Nash equilibrium $\bm\sigma$ such that ${\sf F}({\bm\sigma})\leq 2{\sf F}({\bm\sigma}^*(G))$ when $\kappa$ is even and ${\sf F}({\bm\sigma})={\sf F}({\bm\sigma}^*(G))$ when $\kappa$ is odd. Given the result on the price of stability of Theorem \ref{lbposuniform}, these are the best achievable performance.

\begin{theorem}\label{ubposuniform}
For each $m\geq 2$, $\pos({\cal U}^{even}_{m})\leq 2$ and $\pos({\cal U}^{odd}_{m})=1$. Moreover, for any game $G\in {\cal U}_{m}$, a Nash equilibrium $\bm\sigma$ such that ${\sf F}({\bm\sigma})\leq 2{\sf F}({\bm\sigma}^*(G))$ if $G\in{\cal U}^{even}_{m}$ and such that ${\sf F}({\bm\sigma})= {\sf F}({\bm\sigma}^*(G))$ if $G\in{\cal U}^{odd}_{m}$ can be computed in pseudo-polynomial time.
\end{theorem}
\begin{IEEEproof}
Fix an integer $m\geq 2$ and a game $G\in {\cal U}_{m}$. We prove the claim by showing that Algorithm \ref{alg:uniformne} computes a Nash equilibrium $\bm\sigma$ for $G$ such that ${\sf F}({\bm\sigma})\leq 2{\sf F}({\bm\sigma}^*(G))$ if $G\in{\cal U}^{even}_{m}$ and such that ${\sf F}({\bm\sigma})={\sf F}({\bm\sigma}^*(G))$ if $G\in{\cal U}^{odd}_{m}$.

\alglanguage{pseudocode}
\begin{algorithm*}[h]
\caption{It takes as input a colorful bin packing game with uniform sizes $G$}\label{alg:uniformne}
\begin{algorithmic}[1]
\State {$X\leftarrow \{x_1,\ldots,x_n\}$}
\State {$i\leftarrow 1$}
\State {$c_{old}\leftarrow 0$}
\While {$(X\neq\emptyset)$}
\If {$( |B_i| < \kappa$) \&\& ($\exists x_j\in X$ s.t. $c_j \neq c_{old}) )$}\label{prescelta}
\State {$c\leftarrow$ most frequent color among the items in $X$ having color other than $c_{old}$}\label{scelta}
\State {Select an item $x_j$ of color $c$}
\State {$X\leftarrow X\setminus\{ x_j\}$}
\State {$c_{old}\leftarrow c$}
\State {$\sigma_j\leftarrow B_i$}
\Else
\State {$i\leftarrow i+1$}
\State {$c_{old}\leftarrow 0$}
\EndIf
\EndWhile
\State \Return{$\bm\sigma$}
\end{algorithmic}
\end{algorithm*}

We start by showing that the strategy profile $\bm\sigma$ returned by Algorithm \ref{alg:uniformne} is a Nash equilibrium for $G$. Let us partition $\bm\sigma$ into three sets, namely $\Gamma, \Delta, \Theta$, where $\Gamma$ contains all the full bins, $\Delta$ contains all the non-full and non-singleton bins and $\Theta$ contains all the singleton bins. It is not difficult to see that, by the definition of Algorithm \ref{alg:uniformne}, $\Delta$ and $\Theta$ are such that {\em (i)} $\Delta$ is either empty or contains only one bin, {\em (ii)} all items stored into bins belonging to $\Theta$ have the same color, denoted as $c_{\Theta}$, {\em (iii)} the item on top of the bin in $\Delta$ (if any) has color $c_\Theta$.

Now assume, by way of contradiction, that there exists a player $j$ possessing an improving deviation in $\bm\sigma$ towards a bin $B_i$. Clearly, this can only be possible if $x_j$ is packed into a singleton bin and $B_i\in\Delta\cup\Theta$, but properties ({\em ii}) and {\em (iii)} above imply a contradiction. So, $\bm\sigma$ is a Nash equilibrium.

Let $n_{z}(c)$ be the number of items of color $c$ belonging to $X$ at the $z$th iteration of Algorithm \ref{alg:uniformne}.

\begin{lemma}
If either $|\Theta|\geq 2$ or $|\Delta|=|\Theta|=1$, then the color of each item occupying an odd position in a bin belonging to $\Gamma\cup\Delta$ is $c_\Theta$.
\label{alternato}
\end{lemma}
\begin{IEEEproof}
Consider an iteration $z$ of Algorithm \ref{alg:uniformne} such that $c_{old}\neq c_\Theta$ and an item of color $c\neq c_\Theta$ is selected. As color $c_\Theta$ is a candidate color among the ones considered at line \ref{scelta} of the algorithm, it must be $n_z(c)\geq n_z(c_\Theta)$. For the case of $|\Theta|\geq 2$, let $\overline{z}$ be the first iteration at which the algorithm starts constructing singleton bins, while, for the case of $|\Delta|=|\Theta|=1$, let $\overline{z}$ be the iteration at which the algorithm selects the last item packed into the unique bin in $|\Delta|$. In both cases, it follows that $n_{\overline{z}}(c_\Theta)-n_{\overline{z}}(c)\geq 2$.

Let $z'$ be the first iteration, among the ones realized after iteration $z$, in which an item of color $c_\Theta$ is selected and such that $n_{z'}(c_\Theta)-n_{z'}(c)\geq 2$. Clearly, $z'$ is well-defined because iteration $\overline{z}$ meets the required conditions. This implies that the difference between the number of items of color $c$ and the number of items of color $c_\Theta$ selected by Algorithm \ref{alg:uniformne} during all iterations going from $z$ to $z'-1$ is at least $2$. Hence, there is an iteration $z''$ at which an item of color $c$ is selected despite the fact that $n_{z''}({c_\Theta})= n_{z''}(c)+1$. By line \ref{prescelta} of the algorithm, this can happen only if $c_{old}=c_\Theta$ which implies that, at iteration $z''-1$, an item of color $c_\Theta$ is selected which gives $n_{z''-1}(c_\Theta)=n_{z''}(c_\Theta)+1= n_{z''}(c)+2$ which contradicts the minimality of $z'$.

Hence, we have proved that, at each iteration such that $c_{old}\neq c_\Theta$, Algorithm \ref{alg:uniformne} selects an item of color $c_\Theta$ which implies the claim.
\end{IEEEproof}

By the previous lemma, we get the following corollary which gives us the number of items of color $c_\Theta$ and the number of items of color different that $c_\Theta$.

\begin{corollary}
If either $|\Theta|\geq 2$ or $|\Delta|=|\Theta|=1$, then each bin in $B_j\in O({\bm\sigma})$ contains $\lceil |B_j({\bm\sigma})|/2\rceil$ items of color $c_\Theta$.
\label{alternatocorollario}
\end{corollary}

Let $\#c_\Theta$ be the number of items having color $c_\Theta$. We conclude by showing that ${\sf F}({\bm\sigma})\leq 2{\sf F}({\bm\sigma}^*(G))$ when $\kappa$ is even and that ${\sf F}({\bm\sigma})={\sf F}({\bm\sigma}^*(G))$ when $\kappa$ is odd. Towards this end, we use Corollary \ref{alternatocorollario} together with the simple basic fact.

\begin{fact}\label{basic}
$2\#c_\Theta\leq n+{\sf F}({\bm\sigma}^*(G))$.
\end{fact}

Let us start with the cases not covered by Corollary \ref{alternatocorollario}, that is, $|\Theta|=0$ and $|\Theta|=1\wedge |\Delta|=0$. In both cases, we have ${\sf F}({\bm\sigma})={\sf F}({\bm\sigma}^*(G))$ independently of the parity of $\kappa$, as $O({\bm\sigma})$ contains at most one non-full bin. Thus, in the remaining of the proof, we can assume that Corollary \ref{alternatocorollario} holds. Let $\delta\in\{0,\ldots,\kappa-1\}$ be the number of items stored into the bin belonging to $\Delta$ ($\delta=0$ models the case in which this bin does not exist).

For the case in which $\kappa$ is odd, by Corollary \ref{alternatocorollario}, we have $\#c_\Theta=|\Gamma|\frac{\kappa+1}{2}+\left\lceil\frac{\delta}{2}\right\rceil+|\Theta|$ and $n=|\Gamma|\kappa+\delta+|\Theta|$. Assume, by way of contradiction, that ${\sf F}({\bm\sigma}^*(G))<{\sf F}({\bm\sigma})$, that is, ${\sf F}({\bm\sigma}^*(G))\leq |\Gamma|+|\Delta|+|\Theta|-1$. By Fact \ref{basic}, we obtain
\begin{equation}\label{finaledispari}
2\left\lceil\frac \delta 2\right\rceil\leq\delta+|\Delta|-1.
\end{equation}
Now observe that, for $|\Delta|=0$, which implies $\delta=0$, (\ref{finaledispari}) is not satisfied. Hence, it must be $|\Delta|=1$ which, as $|\Theta| \geq 1$ (recall that we are under the hypothesis in which Corollary \ref{alternatocorollario} holds), implies that $\delta > 1$. Now, if $\delta$ is even, by Corollary \ref{alternatocorollario}, the item on top of the unique bin in $\Delta$ has color different than $c_\Theta$. This means that a player controlling an item packed into any bin in $\Theta$ has an improving deviation by migrating to the unique bin in $\Delta$, thus contradicting the fact that $\bm\sigma$ is a Nash equilibrium. Thus, under the hypothesis of $|\Delta|=1$ and $\delta$ odd, (\ref{finaledispari}) is again not satisfied, thus rising a contradiction. Hence, it follows that ${\sf F}({\bm\sigma}^*(G))={\sf F}({\bm\sigma})$.

For the case in which $\kappa$ is even, by Corollary \ref{alternatocorollario}, we have $\#c_\Theta=|\Gamma|\frac{\kappa}{2}+\left\lceil\frac{\delta}{2}\right\rceil+|\Theta|$ and $n=|\Gamma|\kappa+\delta+|\Theta|$. As ${\sf F}({\bm\sigma}^*(G))\geq |\Gamma|+|\Delta|$, if $|\Gamma|\geq |\Theta|$, it follows ${\sf F}({\bm\sigma})=|\Gamma|+|\Delta|+|\Theta|\leq 2|\Gamma|+|\Delta|\leq 2{\sf F}({\bm\sigma}^*(G))$. Thus, in the remaining of the proof, we assume that $|\Theta|>|\Gamma|$. Assume now, by way of contradiction, that ${\sf F}({\bm\sigma}^*(G))<{\sf F}({\bm\sigma})/2$, which implies ${\sf F}({\bm\sigma}^*(G))\leq\frac{|\Gamma|+|\Delta|+|\Theta|}{2}-\frac 1 2$. By Fact \ref{basic}, we obtain
\begin{equation}\label{finalepari}
2\left\lceil\frac \delta 2\right\rceil+\frac{|\Theta|}{2}\leq\delta+\frac{|\Gamma|+|\Delta|-1}{2}.
\end{equation}
Using the hypothesis that $|\Theta|>|\Gamma|$ within (\ref{finalepari}), we obtain
\begin{equation}
2\left\lceil\frac \delta 2\right\rceil\leq\delta+\frac{|\Delta|}{2}-1
\end{equation}
which is never satisfied, thus rising a contradiction. Hence, it follows that ${\sf F}({\bm\sigma})\leq 2{\sf F}({\bm\sigma}^*(G))$.

We now argue the complexity of Algorithm \ref{alg:uniformne}. We first notice that, for uniform sizes, the compact representation of the input has size $\Omega(m + \log n)$. Moreover, it is easy to see that Algorithm \ref{alg:uniformne} has complexity $O(n)$. It turns out that when, for instance, $m=\Omega(n^{\frac{1}{h}})$, for some constant $h$, the algorithm has polynomial time complexity. However, when $m=O(\log n)$, the complexity is pseudo-polynomial.   
\end{IEEEproof}

Theorems \ref{lbposuniform} and \ref{ubposuniform} completely characterize the $\pos$ of colorful bin packing games with uniform sizes. For what concerns the $\poa$, we also obtain a complete picture by means of the following results.

For the case of at least three colors, the $\poa$ can be arbitrarily high.

\begin{theorem}\label{POA_unbounded_uniform_m>=3}
For each $m\geq 3$, both $\poa({\cal U}^{odd}_{m})$ and $\poa({\cal U}^{even}_{m})$ are unbounded.
\end{theorem}

\begin{IEEEproof}
We show that there exist two colorful bin packing games $G\in {\cal U}^{even}_{3}$ and $G'\in {\cal U}^{odd}_{3}$ whose $\poa$ grows asymptotically with the number of their players. It is easy to adapt the proof so as to deal with any number of colors $m \geq 3$.
Game $G$ is as follows:
there are $n=4k$ players, all of them of size $\epsilon$, where $k\geq 1$ is an arbitrary integer, and $\epsilon>0$ is an arbitrary small real value such that $\sum_{i=1}^{n}s_i\leq 1$, so that $\kappa=4k$ is even. There are $2k$ items of color $c_1$ denoted as $b_1, \ldots ,b_{2k}$, $k$ items of color $c_2$ denoted as $w_1, \ldots, w_k$, and $k$ items of color $c_3$ denoted as $r_1, \ldots, r_k$.
Notice that ${\sf F}({\bm\sigma}^*)=1$, since all items can be feasibly packed into the same bin, as depicted in Figure \ref{fig3} on the left side.
In the same figure, on the right side, is also depicted a packing corresponding to a Nash equilibrium ${\bm\sigma}\in{\sf NE}(G)$ with social cost
${\sf F}({\bm\sigma})=2k=\frac{n}{2}$, which yields $\poa(G)=\Omega(n)$.

Game $G'$ can be obtained by adding item $b_0$ of color $c_1$ to game $G$. This item can be feasibly packed at the bottom of both bins $B_1$ depicted in Figure \ref{fig3}, so that we get $\kappa=4k+1$, which is odd, and $\poa(G')=\Omega(n)$.
\begin{figure}[ht]
\centering
\includegraphics[scale=0.76]{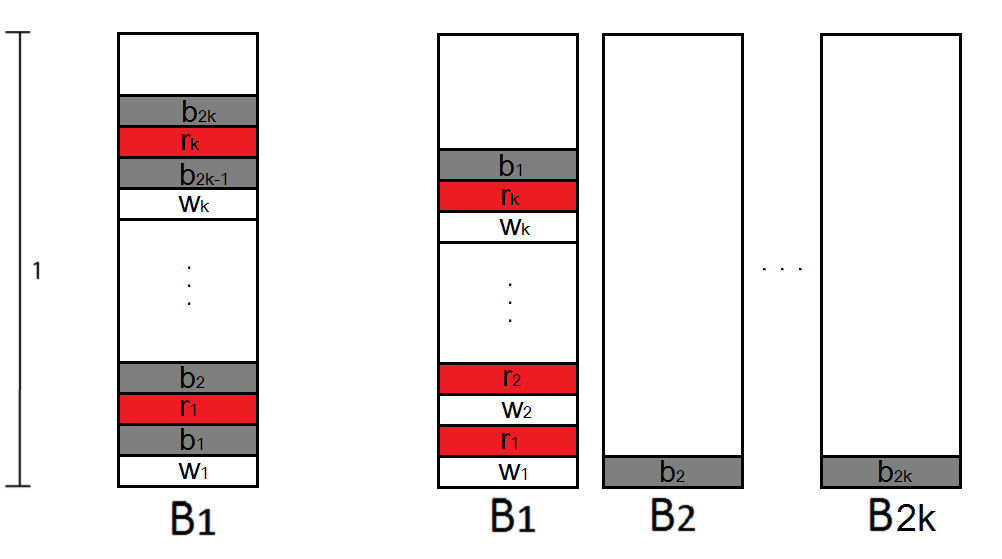}
\caption{A colorful bin packing game $G\in {\cal U}_{3}$ with $\poa(G) \geq \frac{n}{2}$:
on the left side the packing corresponding to a social optimum and on the right side the packing corresponding to a Nash equilibrium with social cost $2k$.}
\label{fig3}
\end{figure}
\end{IEEEproof}

For the case of black and white bin packing games, we show a lower bound of $3$ on the $\poa$ of games for which $\kappa$ is an odd number, thus matching the upper bound showed in Theorem \ref{ubpoa} which holds for general sizes.

\begin{theorem}\label{lowerbound_poa_uniform_k_odd_two_colors}
$\poa({\cal U}^{odd}_2)\geq 3$.
\end{theorem}

\begin{IEEEproof}
We prove the claim by showing that, for any fixed $\epsilon>0$, there exists a game $G_{\epsilon}\in{\cal U}^{odd}_2$ such that $\poa(G_\epsilon)\geq 3-\epsilon$.

Game $G_{\epsilon}$ is defined as follows: there are $n=k(k+3)/2$ items, of which $(k^2+4k-1)/4$ are white and the remaining $(k+1)^2/4$ are black, where $k$ is an odd number such that $k\geq (8-3\epsilon)/\epsilon$. The size of each item is set in such a way that $\kappa = k$.

Let ${\bm\sigma}^*$ be the strategy profile such that $O({\bm\sigma}^*)=(B_1,\ldots,B_{(k+1)/2+1})$ where
\begin{itemize}
\item[$\bullet$] each of the first $(k+1)/2$ bins contains $(k+1)/2$ white items and $(k-1)/2$ black items;
\item[$\bullet$] bin $B_{(k+1)/2+1}$ contains $(k-1)/2$ white items and $(k+1)/2$ black items.
\end{itemize}
Since, by definition, at most $k$ items can be packed into a bin and $k$ is odd, we have ${\sf F}({\bm\sigma}^*(G_{\epsilon}))\geq\left\lceil\frac n k\right\rceil=\left\lceil\frac{k+3}{2}\right\rceil=(k+1)/2+1$, so that ${\bm\sigma}^*$ is a social optimum.

Let ${\bm\sigma}$ be the strategy profile such that $O({\bm\sigma}^*)=(B_1,\ldots,B_{(3k+1)/2})$ where
\begin{itemize}
\item[$\bullet$] each of the first $(k+1)/2$ bins contains $(k-1)/2$ white items and $(k+1)/2$ black items;
\item[$\bullet$] each of the remaining $k$ bins contains a single white item.
\end{itemize}
Since all of the first $(k+1)/2$ bins are full and the remaining $k$ bins are singletons and contain only white items, it follows that $\bm\sigma$ is a Nash equilibrium. By the definition of $k$, we get $\poa(G_\epsilon)\geq\frac{3k+1}{k+3}\geq 3-\epsilon$.
\end{IEEEproof}

For the leftover case of black and white bin packing games for which $\kappa$ is even, we show that the upper bound on the $\poa$ drops to $2$ which matches the lower bound given in Theorem \ref{lbposuniform} for the $\pos$.

\begin{theorem}\label{upperbound_poa_twocolors_k_even}
$\poa({\cal U}^{even}_2)\leq 2$.
\end{theorem}
\begin{IEEEproof}
Fix a game $G\in{\cal U}^{even}_2$ and a Nash equilibrium $\bm\sigma$ for $G$. Let us partition $\bm\sigma$ into three sets, namely $\Gamma, \Delta, \Theta$, where $\Gamma$ contains all the full bins, $\Delta$ contains all the non-full and non-singleton bins and $\Theta$ contains all the singleton bins. It is not difficult to see that, by the fact that $\bm\sigma$ is a Nash equilibrium, $\Delta$ and $\Theta$ are such that {\em (i)} $\Delta$ is either empty or contains only one bin, {\em (ii)} all items stored into bins belonging to $\Theta$ have the same color, which we assume without loss of generality to be black, {\em (iii)} the item on top of the bin in $\Delta$ (if any) is black. Let $\delta\in\{0,\ldots,\kappa-1\}$ be the number of items packed into the unique bin in $\Delta$ ($\delta=0$ model the case in which $\Delta=\emptyset$). Since $\kappa$ is even and the item on top of the bin in $\Delta$ (if any) is black, we have
$\#B = |\Gamma|\frac\kappa 2+\left\lceil\frac\delta 2\right\rceil+|\Theta|$ and $\#W = |\Gamma|\frac\kappa 2+\left\lfloor\frac\delta 2\right\rfloor$. By substituting the values within inequality (\ref{eq1}), we obtain
$$|\Gamma|\frac\kappa 2+\left\lceil\frac\delta 2\right\rceil+|\Theta|\leq|\Gamma|\frac\kappa 2+\left\lfloor\frac\delta 2\right\rfloor+{\sf F}({\bm\sigma}^*(G))$$ which implies ${\sf F}({\bm\sigma}^*(G))\geq |\Theta|$. Moreover, as all bins in $\Gamma$ is full, we have ${\sf F}({\bm\sigma}^*(G))\geq |\Gamma|+|\Delta|$, so that ${\sf F}({\bm\sigma}^*(G))\geq \max\{|\Theta|,|\Gamma|+|\Delta|\}$.  By the arbitrariness of $\bm\sigma$, we obtain $$\poa(G)=\frac{{\sf F}({\bm\sigma})}{{\sf F}({\bm\sigma}^*(G))}\leq\frac{|\Gamma|+|\Delta|+|\Theta|}{\max\{|\Theta|,|\Gamma|+|\Delta|\}}\leq 2.$$ 
\end{IEEEproof}


%


\begin{thebibliography}{99}

\bibitem{AE13}
R. Adar, L. Epstein. Selfish bin packing with cardinality constraints. {\em Theoretical Computer Science}, 495:66-80, 2013.




\bibitem{BBDEKT15}
J. Balogh, J. B{\'{e}}k{\'{e}}si, G. D{\'{o}}sa, L. Epstein, H. Kellerer, Z. Tuza. Online Results for Black and White Bin Packing. {\em Theory of Computing Systems}, 56(1):137-155, 2015.

\bibitem{BBDEKLT15}
J. Balogh, J. B{\'{e}}k{\'{e}}si, G. D{\'{o}}sa, L. Epstein, H. Kellerer, A. Levin, Z. Tuza. Offline black and white bin packing. {\em Theoretical Computer Science}, 596:92-101, 2015.

\bibitem{B06}
V. Bil{\`{o}}. On the packing of selfish items. In {\em Proceedings of the 20th International Parallel Symposium on Distributed Processing (IPDPS)}, 2006.


\bibitem{BSV14}
M. B{\"{o}}hm, J. Sgall, P. Vesel{\'{y}}. Online Colored Bin Packing. In {\em Proceedings of the 12th International Workshop on Approximation and Online Algorithms (WAOA)}, pages 35-46, 2014.

\bibitem{BDESV}
M. B{\"{o}}hm, G. D{\'{o}}sa, L. Epstein, J. Sgall, P. Veselý. Colored Bin Packing: Online Algorithms and Lower Bounds. {\em Algorithmica}, to appear.

\bibitem{CY11}
Z. Cao, X. Yang. Selfish Bin Covering. {\em Theoretical Computer Science}, 412(50):7049-7058, 2011.

\bibitem{CGJ96}
E. G. Coffman Jr., M. R. Garey, D. S. Johnson. Approximation algorithms for bin packing: a survey. Chapter 2 In, Approximation algorithms for NP-hard problems, PWS Publishing Co, 1996.

\bibitem{DE14}
G. D{\'{o}}sa, L. Epstein. The Convergence Time for Selfish Bin Packing. In {\em Proceedings of the 7th International Symposium on Algorithmic Game Theory (SAGT)}, pages 37-48, 2014.

\bibitem{DE12}
G. D{\'{o}}sa, L. Epstein. Generalized selfish bin packing. CoRR, abs/1202.4080 (2012)

\bibitem{DE14b}
G. D{\'{o}}sa, L. Epstein. Colorful Bin Packing. In {\em Proceedings of the 14th Scandinavian Symposium and Workshops on Algorithm Theory (SWAT)}, pages 170-181, 2014.

\bibitem{E13}
L. Epstein. Bin Packing Games with Selfish Items. In {\em Proceedings of the 38th International Symposium on Mathematical Foundations of Computer Science (MFCS)}, pages 26-30, 2013.

\bibitem{EK11}
L. Epstein, E. Kleiman. Selfish Bin Packing. {\em Algorithmica}, 60(2):368-394, 2011.

\bibitem{EK15}
L. Epstein, E. Kleiman. Selfish Vector Packing. In {\em Proceedings of the 23rd European Symposium on Algorithms (ESA)}, pages 471-482, 2015.

\bibitem{EKLS11}
L. Epstein, S. O. Krumke, A. Levin, H. Sperber. Selfish bin coloring. {\em Journal of Combinatorial Optimization}, 22(4):531-548, 2011.

\bibitem{FFMW11}
C. G. Fernandes, C. E. Ferreira, F. K. Miyazawa, Y. Wakabayashi. Selfish Square Packing. {\em Elec. Notes in Discrete Math.}, 37(1):369-374, 2011.

\bibitem{KPRS08}
S. O. Krumke, W. de Paepe, J. Rambau, L. Stougie. Bincoloring. {\em Theoretical Computer Science}, 407(1-3):231-241, 2008.

\bibitem{MDHTYZ13}
R. Ma, G. D{\'{o}}sa, X. Han, H. Ting, D. Ye, Y. Zhang. A note on a selfish bin packing problem. {\em Journal of Global Optimization}, 56(4):1457-1462, 2013.

\bibitem{YZ08}
G. Yu, G. Zhang. Bin Packing of Selfish Items. In {\em Proceedings of the 4th International Workshop on Internet and Network Economics (WINE)}, pages 446-453, 2008.

\end{thebibliography}
\end{document}